\documentclass[fleqn,usenatbib]{mnras}

\usepackage{newtxtext,newtxmath}

\usepackage[T1]{fontenc}
\usepackage{ae,aecompl}

\usepackage{graphicx}	
\usepackage{amsmath}	

\usepackage{chemformula}
\usepackage{threeparttable}
\usepackage{longtable}
\usepackage{bibentry}
\usepackage{eucal}
\usepackage{siunitx}
\usepackage{anyfontsize}

\usepackage{longtable}
\usepackage{pdflscape}
\usepackage[version=4]{mhchem}
\usepackage{xr-hyper}

\title[White dwarf constraints on geology]{White dwarf constraints on geological processes at the population level}

\author[A. M. Buchan et al.]{
Andrew M. Buchan$^{1,2}$, Amy Bonsor$^{2}$, Laura K. Rogers$^{2}$, Marc G. Brouwers$^{2}$, \newauthor Oliver Shorttle$^{2,3}$, Pier-Emmanuel Tremblay$^{1}$\thanks{E-mail: andy.buchan@warwick.ac.uk (AMB)}
\\
$^{1}$Department of Physics, University of Warwick, Coventry CV4 7AL, UK\\
$^{2}$Institute of Astronomy, University of Cambridge, Madingley Road, Cambridge CB3 0HA, UK\\
$^{3}$Department of Earth Sciences, University of Cambridge, Downing Street, Cambridge CB2 3EQ, UK\\
}

\date{Accepted XXX. Received YYY; in original form ZZZ}

\pubyear{2023}

\begin{document}
\label{firstpage}
\pagerange{\pageref{firstpage}--\pageref{lastpage}}
\maketitle

\begin{abstract}
White dwarf atmospheres are frequently polluted by material from their own planetary systems. Absorption features from Ca, Mg, Fe and other elements can provide unique insights into the provenance of this exoplanetary material, with their relative abundances being used to infer accretion of material with core- or mantle-like composition. Across the population of white dwarfs, the distribution of compositions reveals the prevalence of geological and collisional processing across exoplanetary systems. By predicting the distribution of compositions in three evolutionary scenarios, this work assesses whether they can explain current observations. We consider evolution in an asteroid belt analog, in which collisions between planetary bodies that formed an iron core lead to core- or mantle-rich fragments. We also consider layer-by-layer accretion of individual bodies, such that the apparent composition of atmospheric pollution changes during the accretion of a single body. Finally, we consider that compositional spread is due to random noise. We find that the distribution of Ca, Fe and Mg in a sample of 202 cool DZs is consistent with the random noise scenario, although 7 individual systems show strong evidence of core-mantle differentiation from additional elements and/or low noise levels. Future surveys which detect multiple elements in each of a few hundred white dwarfs, with well understood biases, have the potential to confidently distinguish between the three models.

\end{abstract}




\section{Introduction}

The ruins of ancient planetary systems pollute the atmospheres of many white dwarfs. Between 27\% and 50\% of young white dwarfs have heavy elements in their atmospheres \citep{Zuckerman2010,Koester2014}. The detection of such elements is typically attributed to the recent accretion of planetary material that has survived into the post-main sequence phase of its host star (e.g., \citealt{Jura2003,Zuckerman2010,JuraYoung2014}). White dwarfs polluted by such material offer a unique opportunity to examine the composition of planetary building blocks, and hence to understand the key processes which govern their formation and evolution. A key process is core--mantle differentiation: the formation of a Fe-rich core and an Fe-poor mantle.

The study of metal abundances in white dwarfs has focussed on modelling systems one at a time, revealing geological histories at the individual level. For example, the Fe-rich pollution of \mbox{PG\;0843+516} is suggestive of a body which underwent core--mantle differentiation and then suffered a catastrophic loss of much of its mantle \citep{Gaensicke2012,Xu2019}. Other individual polluted white dwarfs record the accretion of water-rich bodies (e.g., \citealt{Farihi2013}), crustal material (e.g., \citealt{Zuckerman2011}), material which lost volatile elements during a magma ocean phase \citep{Harrison2021b} and material potentially derived from icy exomoons \citep{Klein2021,Doyle2021}.

Individual systems can provide insight into which evolutionary pathways may operate in exoplanetary systems, but in order to determine the relative importance of these pathways, population level analysis is necessary. This work performs such a population level analysis, and shows that it is already possible to extract results from current white dwarf samples. The best known samples with relevance to metal pollution are the cool DZ sample \citep{Hollands2017}, the He-rich population of \citet{Coutu2019} and the 40\;pc \citep{Tremblay2020,McCleery2020,OBrien2023,OBrien2024} and 100\;pc samples \citep{JimenezEsteban2018,GentileFusillo2018,Kilic2020}. Ongoing surveys such as the Sloan Digital Sky Survey (SDSS) and the Dark Energy Spectroscopic Instrument (DESI) continue to identify new polluted white dwarfs \citep{Manser2024DESI}. Such ground-based spectroscopic follow up of white dwarfs identified with \textit{Gaia} should yield more than 1,000 systems amenable to detailed abundance analysis, with the upcoming 4MOST and WEAVE-WD surveys providing spectroscopy for upwards of 100,000 white dwarfs \citep{Chiappini2019,Gaensicke2019WhitePaper}. This will dramatically increase the sample sizes available to population studies.

Such large populations of polluted white dwarfs will allow us to learn about the physical processes which are generally important across all exoplanetary systems, the signatures of which may be identifiable in sufficiently large samples. The composition of planetary material in the atmospheres of white dwarfs can tell us about the geological process of iron core formation, but only if material from the iron-core and silicate mantle are separated. This separation can occur due to violent collisions that break-up the planetary bodies into small fragments or during the accretion process itself, where bodies may be accreted layer-by-layer.

Collisional evolution is a key process in the early Solar System. The composition of terrestrial planets can be altered if they suffer sufficient collisional processing \citep{Marcus2009,Stewart2012,Carter2015}, a scenario which may explain Mercury's high density \citep{Benz1988,Benz2007} and the formation of the Moon \citep{Hartmann1975,Cameron1976}. Collisional evolution has shaped the size distribution of the asteroid belt and Kuiper belt \citep{Bottke2005,Kenyon2004,Holsapple2022}. Similarly, debris discs in other planetary systems are governed by collisional processes \citep{Krivov2006,Thebault2007}, as evidenced by ALMA observations \citep{Marino2021,ImazBlanco2023}. Bodies from these collisionally evolved structures may survive into the post-main sequence phase, ultimately supplying the central white dwarf with pollution \citep{Bonsor2011}.

The exact manner in which the accretion of planetary bodies onto white dwarfs proceeds remains poorly understood. Any process in which a body is not fully accreted at once, but instead is accreted layer-by-layer over multiple sinking timescales, has the potential to separate the core and mantle material. One such process was suggested by \citet{Brouwers2023a}, and is referred to in this work as `orbit-by-orbit' accretion. Orbit-by-orbit accretion is based on the commonly suggested scenario that accreted material is derived from a tidally disrupted asteroid \citep{Debes2002,Jura2003,Zuckerman2007,Klein2010,Dufour2010,Veras2014,Malamud2020b,Malamud2021,Veras2021}. Material from different locations in the asteroid may be spread across different orbits according to orbital energy, a process that is also seen in SPH simulations \citep{Malamud2020a}. This material may then be accreted orbit-by-orbit. If the disrupted asteroid was core--mantle differentiated, this leads to asynchronous accretion of core and mantle.

Collisional evolution and orbit-by-orbit accretion predict different distributions for the fraction of accreted material which is core-like, meaning that these processes could be distinguished given a sufficiently large sample of polluted white dwarfs. This would provide valuable insight into which processes are important in driving the evolution of pollutants.

The present study predicts the number of white dwarfs required to place constraints on the process(es) dominating the evolutionary history of pollutants. In order to make this prediction, we outline a method in Section~\ref{sec:synthetic_methods} that synthesises a population of polluted white dwarfs. Our synthesis method takes into account random noise, and also detection bias caused by the inability to detect metals which are only present in small quantities. In Section~\ref{sec:synthetic_results}, we use this pipeline to identify the impact of these biases on our interpretation of the population of exoplanetary bodies accreted by white dwarfs. We also address the important question of how many polluted white dwarfs are necessary to distinguish samples in which the collisional and orbit-by-orbit scenarios are dominant (when the underlying pollutant population is differentiated), and also to distinguish these cases from a control case in which pollutants are not core--mantle differentiated. We then assess whether this can already be achieved for the cool DZ sample of \citet{Hollands2017}. We discuss the implications for the field in Section~\ref{sec:synthetic_discussion} and summarise in Section~\ref{sec:synthetic_conclusions}.

\section{Methods}
\label{sec:synthetic_methods}

Populations of polluted white dwarfs can reveal the underlying mechanisms that control the evolution of exoplanetary bodies, as long as we can relate the observed compositional distribution to the true distribution. We focus on trends in the core- or mantle-like nature of accreted material, and relate these trends to those predicted by three different evolutionary processes.

\begin{figure*}
    \centering
    \includegraphics[width=\textwidth,keepaspectratio=true]{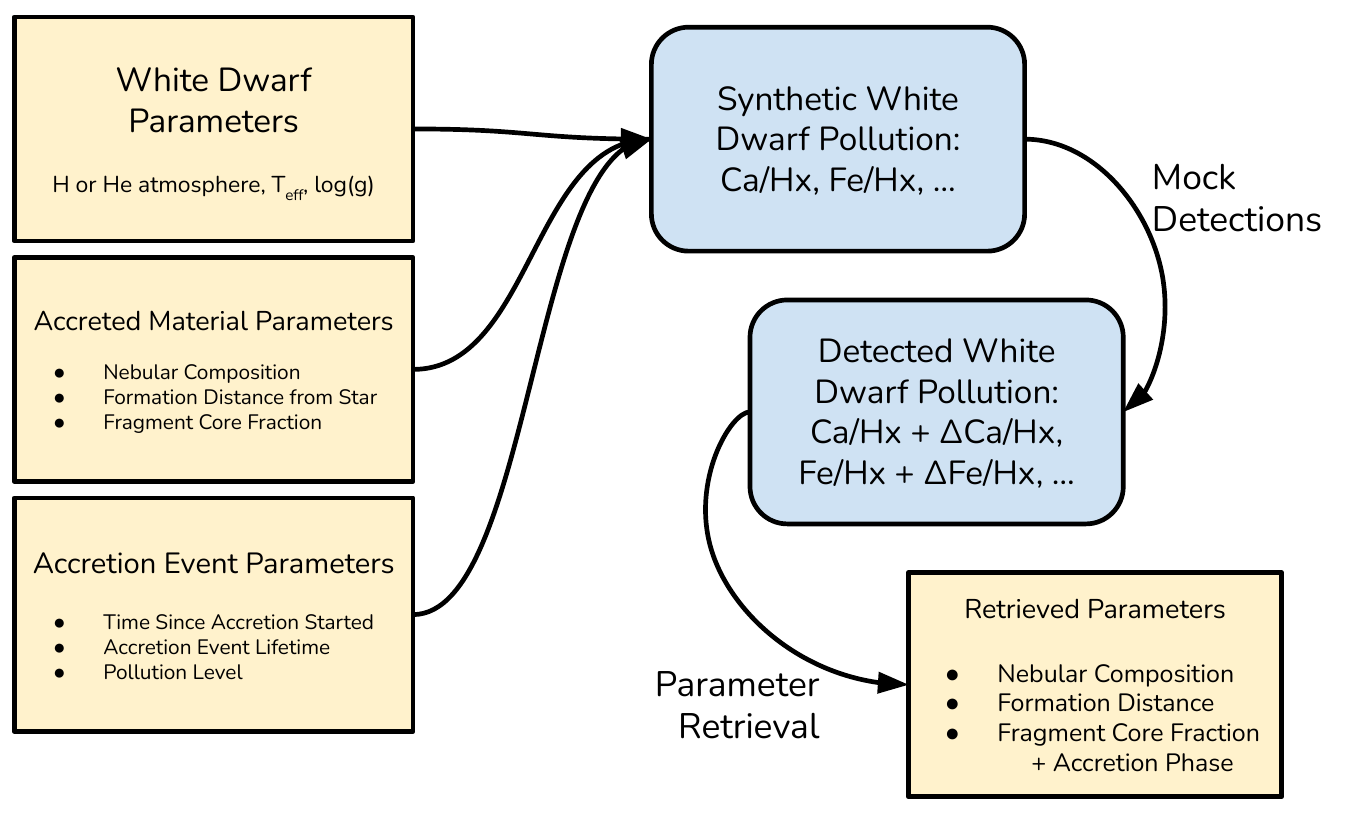}
    \caption[Flowchart of the white dwarf pipeline]{Flowchart of the pipeline for a single synthetic white dwarf as described in Section~\ref{sec:synthetic_methods}. The first stage is the random generation of the various parameters describing the white dwarf itself, the accreted material, and the accretion process. From this, we calculate the abundances of elements including Ca and Fe in the white dwarf's atmosphere (see Section~\ref{sec:synthetic_generation}). Hx refers to the dominant element in the atmosphere, either H or He. We then calculate mock elemental detections for the resulting white dwarf (see Section~\ref{sec:synthetic_observation}). Detection thresholds and random noise are applied, resulting in a set of detected elemental abundances which, in general, differ from those calculated in the previous step. Elements can also fail to be detected. This process introduces systematic bias. Finally, we use a simple algorithm (see Section~\ref{sec:synthetic_modelling}) for parameter retrieval. This is repeated for a large population of synthetic white dwarfs. Statistical trends extracted from the population deviate from the input prescriptions due to observational bias, random noise, and imperfect modelling.}
\label{fig:pipeline}
\end{figure*}

We aim to compare a real sample of white dwarfs against a synthetic sample, and test them for consistency. To do this, we introduce a pipeline which generates synthetic white dwarfs. The stages of this pipeline are illustrated in Figure~\ref{fig:pipeline}, and described in further detail in the following subsections.

\subsection{Generating a synthetic population}
\label{sec:synthetic_generation}

\begin{table}
\centering
\caption[Synthetic white dwarf parameters]{The parameters used to generate synthetic populations of white dwarfs with H- and He-dominated atmospheres. A uniform fragment core number fraction distribution is also used in Sections~\ref{sec:control}, \ref{sec:controlrealistic} and \ref{sec:controlrealisticnoise} for testing purposes only.}
\label{tab:synthetic_parameters}
\begin{tabular}{ccc}
\hline
Parameter & \multicolumn{2}{c}{Distribution}\\ 
 & H-dominated & He-dominated\\ 
\hline
${\rm T}_{\textrm{eff}}$ & 40 pc DAs & \citet{Hollands2017} \\
$\log(g)$ & 40 pc DAs & \citet{Hollands2017} \\
Initial Nebular Composition & \multicolumn{2}{c}{Uniform across 958 stars} \\
Pollutant Formation Distance & \multicolumn{2}{c}{Log-uniform, $10^{-0.55}$-$10^{-0.25}$ AU} \\
Fragment Core Fraction, $f_{\rm c}$ & \multicolumn{2}{c}{Collisional, Orbit-by-orbit or Delta} \\
Time Since Accretion, $t$ & N/A & Uniform, 0-20\;Myr\\
Accretion Event Lifetime, $t_{\rm event}$ & N/A & Uniform, 0-10\;Myr \\ Pollution Level, $\lambda$ & \multicolumn{2}{c}{Equation~\ref{eq:pol_frac}}\\
\hline
\end{tabular}
\end{table}

Our aim is to produce a population of white dwarfs with a range of metal abundances in their atmospheres at the time of observation. The metal abundances result from the accretion of a realistic range of planetesimals, accreted by white dwarfs in different exoplanetary systems. In order to do this, we use a simple framework to predict the composition of typical planetesimals, and assume that white dwarfs sample from these at random. We consider only variations in the dominant mechanisms controlling planetesimal composition and structure: initial nebular composition and core--mantle differentiation. We ignore the volatile component of planetesimals for the purpose of this work.

A sample of synthetic polluted white dwarfs is generated by randomly sampling the parameters that describe the white dwarf, the material it has accreted, and the timing of the accretion event itself. We then calculate the resulting atmospheric abundances of metals including Ca, Fe, Mg, Al, Ti, Ni and Cr, which are the elements used in this analysis. The forward model used is the same as that of \citet{Harrison2018,Harrison2021}, with updated sinking timescales and crustal material omitted. In this model, the composition of core- and mantle-like material is based on that of Earth, but with relative metal abundances scaled according to nebular composition (and, close to the star, volatile depletion). The variables we sample are as follows.

\subsubsection{White dwarf parameters}

For the white dwarf, the quantities of interest are the effective temperature, $\textrm{T}_{\rm eff}$, the surface gravity, $\log(g)$, and the dominant component of the atmosphere (H or He, which we refer to as Hx for convenience). These variables control the timescales over which elements sink in the white dwarf's atmosphere.

For H-dominated white dwarfs, the values of $\textrm{T}_{\rm eff}$ and $\log(g)$ are randomly sampled from a distribution generated from $\textrm{T}_{\rm eff}$/$\log(g)$ values from all DAs in the 40pc sample \citep{Tremblay2020,McCleery2020,OBrien2023,OBrien2024} as recorded in the Montreal White Dwarf Database\footnote{\url{https://www.montrealwhitedwarfdatabase.org/}, accessed 01/08/2022} (MWDD, \citealt{MWDD}). These values are placed into bins (of width \SI{1800}{K} or 0.16 dex respectively), allowing the probability distribution to be estimated and interpolated. The resulting synthetic ${\rm T}_{\rm eff}$ and $\log(g)$ distributions closely mimic the real distributions without directly replicating any particular real system. We choose the 40\,pc sample as it is almost volume complete ($\approx$96\% at time of data extraction), and so is as unbiased as possible. We use the default values of $\textrm{T}_{\rm eff}$ and $\log(g)$ in the MWDD, which may differ from any given individual source for any given system, but we only require representative distributions so do not expect this to affect our results significantly.

For He-dominated systems, we repeat this process with distributions based on the cool DZ sample of \citet{Hollands2017}, using updated values of $\textrm{T}_{\rm eff}$ and $\log(g)$ found by \citet{Blouin2020} with bin widths of \SI{500}{K} and 0.2 dex.

\subsubsection{Initial nebular composition}
\label{sec:initialcomp}

We assume that pollutants form from the same material as their host star, such that their initial composition matches the initial nebular composition. We use a sample of nearby F, G and K-type stars from \citet{Brewer2016} as a proxy for the probable range of compositions, and assume each of these 958 compositions is equally likely.

\subsubsection{Volatile depletion/Formation distance}

We reduce the quantity of volatile elements in the initial nebular composition according to the ambient temperature during formation in the protoplanetary disc. This process is included for the sake of mimicking the model of \citet{Harrison2018,Harrison2021}, but has minimal effect for the purposes of this paper, so we do not discuss it further.

\subsubsection{Fragment core fraction}

Fragment core fraction ($f_{\rm c}$) is the key variable representing core--mantle differentiation. It accounts for the possibility that any given accreted body is actually a fragment from a larger parent body which has differentiated into a separate core and mantle during formation. Unlike the parent body, the fragment can be composed of an arbitrary combination of core and mantle. $f_{\rm c}$ refers to the core fraction of this accreted fragment, quantified in terms of number (i.e., moles) rather than by weight, and can range from 0 to 1. Higher values of $f_{\rm c}$ cause enrichment in siderophile elements (Fe, Ni, Cr) and depletion in lithophile elements (Ca, Mg, Al, Ti, Na, O).

We do not necessarily assume that core formation always occurs. The signature of pristine, undifferentiated material can be replicated by mixing the core and mantle components in an Earth-like proportion (with $f_{\rm c}\sim0.17$).

The distribution of $f_{\rm c}$ is a key variable in this work, and is described in more detail in Section \ref{sec:fcnf_dists}.

\subsubsection{Accretion phase}

The phase of accretion is crucial because the differential sinking of elements in white dwarf atmospheres distorts what is observed from what was actually accreted, and this effect depends on when the system is observed. We use two variables to describe the accretion phase: the time since accretion began, $t$, and the accretion event lifetime, $t_{\rm event}$. Accretion starts at $t = 0$, and proceeds through three phases. During the first phase (called the `build-up' or `increasing' phase), the composition of atmospheric pollution closely matches the accreted material. If accretion proceeds for a few sinking timescales, the system reaches `steady state', in which all elements are in accretion-diffusion equilibrium. In the steady state phase, elements which sink rapidly through the atmosphere are artificially depleted relative to other elements. Accretion ends at $t = t_{\rm event}$, after which point the system is in the `declining' or `decreasing' phase. If the system is observed in this phase ($t > t_{\rm event}$), the effects of differential sinking can become highly exaggerated.

For H-dominated white dwarfs, sinking timescales are typically short enough that we can safely assume steady state accretion. We therefore fix the parameters of $t$ and $t_{\rm event}$ to arbitrary values corresponding to steady state.

For He-dominated white dwarfs, we assume that $t$ is uniformly distributed between 0-20\;Myr, and that $t_{\rm event}$ is uniformly distributed between 0-10\;Myr. The possible range of $t_{\rm event}$ is not well constrained in the literature. \citet{VerasHeng2020} find that this timescale can range from less than a year to about \SI{1}{Myr}, depending on the parameters of their disc models. Other estimates range from 20\;yr \citep{Wyatt2014} to hundreds of kyr (e.g, \citealt{Rafikov2011b,Girven2012,BuchanThesis}) to $\sim$Myr timescales (e.g., \citealt{Cunningham2021}). Our range spans these possibilities. Our $t$ distribution approximates the timescales over which pollution could plausibly be detected, given our $t_{\rm event}$ distribution.

\subsubsection{Pollution level}

The pollution level, $\lambda$, acts as a proxy for the total mass of accreted material. We define it as the quantity of atmospheric pollution relative to Hx, expressed on a log (base 10) scale such that it is always negative. Systems which have highly negative values of $\lambda$ are only lightly polluted, and their metal abundances may be too low to be detected. Systems with higher (i.e., less negative) values of $\lambda$ are assumed to be rarer. The true $\lambda$ distribution is unknown, so we assume a simple linear distribution $P(\lambda) \propto \lambda + c$, where $P$ is the probability density and $c$ is a constant. We introduce two free parameters, $a$ and $b$, which represent the minimum and maximum allowed values of $\lambda$ respectively. The requirement that $P(\lambda)$ is normalised and continuous at $\lambda = b$ gives $P(\lambda)$ in terms of $a$ and $b$:

\begin{equation}
    P(\lambda) = 
\begin{cases}
    \frac{2(b-\lambda)}{(a-b)^2} ,& \text{if } a \leq \lambda \leq b\\
    0,              & \text{otherwise.}
\end{cases}   
\label{eq:pol_frac}
\end{equation}

We set $a = -12$. This value is chosen primarily to avoid simulating large numbers of systems whose pollution cannot be detected, and has limited physical meaning. We set $b = -3$ (roughly equal to the most polluted white dwarfs), except when mimicking the cool DZs, in which case we calibrate the value of $b$ to $-5$.

For He-dominated systems, the pollution level in principle affects the sinking timescales, which in turn affects the pollution level. Finding a self-consistent solution proved non-trivial. We neglect this effect because, relative to ${\rm T}_{\rm eff}$ and $\log(g)$, pollution level only alters relative sinking timescales by a small amount.

The pollution level, $\lambda$, is sampled independently from $t$ and $t_{\rm event}$, so it is possible for the pollution level to be large while also having $t > t_{\rm event}$. Physically, this means that the system is in the declining phase, and yet there is still a high level of pollution, potentially implying unrealistically large amounts of accreted material. We correct these cases by reducing the pollution level according to the expected exponential loss of material which follows the end of accretion. We calculate a correction factor equal to $\frac{t - t_{\rm event}}{ln(10)\tau_{Ca}}$, where $\tau_{Ca}$ is the sinking timescale of Ca, and subtract this from the pollution level. This correction factor is applied to all white dwarfs in the declining phase of accretion, but only has a significant influence on those with high accretion rates long after accretion has finished.

\subsection{The distribution of fragment core fraction}
\label{sec:fcnf_dists}

This work is based on the key premise that the distribution of core fractions, $f_{\rm c}$, of the fragments found in the atmospheres of white dwarfs differs according to the evolution of these fragments prior to accretion. We consider three different evolutionary models, and their corresponding $f_{\rm c}$ distributions, which are illustrated in Figure~\ref{fig:fcnf_dists}.

\subsubsection{Collisional}

The `Collisional' distribution represents the model in which collisional evolution dominates the evolution of an asteroid or planetesimal belt. As core material is generally stronger and found at greater depth than mantle material, collisional evolution affects this material differently. Gentle collisions chip away at mantles, whilst highly disruptive collisions lead to the merger of iron cores \citep{Marcus2009}. Thus, a population of planetary bodies that start with the same initial $f_{\rm c}$, will, after many years of collisional evolution, contain planetesimals with a wide range of core fractions (e.g., \citealt{Carter2015,Bonsor2020}).

For this work, the collisional distribution is taken from N-body simulations of 100,000 planetesimals as presented by \citet{Bonsor2020}, based on \citet{Carter2015}. Each planetesimal has an initial core mass fraction of 0.35 (corresponding to $f_{\rm c} \approx 0.18$), but as planetesimals collide they produce fragments with differing core fractions. The N-body simulations include prescriptions for multiple collision types, including hit and run, perfect mergers, and partial accretion. The fate of core or mantle material is tracked following each collision. Material that falls below the resolution limit (which is primarily mantle) is assumed to accrete as dust on to the larger bodies. The simulations in \citet{Carter2015} are highly collisional, with multiple collisions per body. The final distribution has a single peak near the initial $f_{\rm c}$, but with significant deviation due to collisional processing.

\subsubsection{Orbit-by-orbit}

A differentiated body passing within the Roche radius of the white dwarf experiences tidal disruption and forms a disc. The disrupted body is unlikely to be accreted within a single sinking timescale, but rather over a significantly longer time period. Following \citet{Brouwers2023a}, we consider the accretion of a single disrupted planetary body to start with the innermost orbit of the tidal disc, before proceeding `orbit-by-orbit' to the outermost orbit. The Orbit-by-orbit distribution mimics the accretion of this disrupted material on the assumption that, at any moment, we don't know which orbit the white dwarf is sampling material from, so it is effectively random.

\begin{figure}
    \centering
    \includegraphics[width=\columnwidth,keepaspectratio=true]{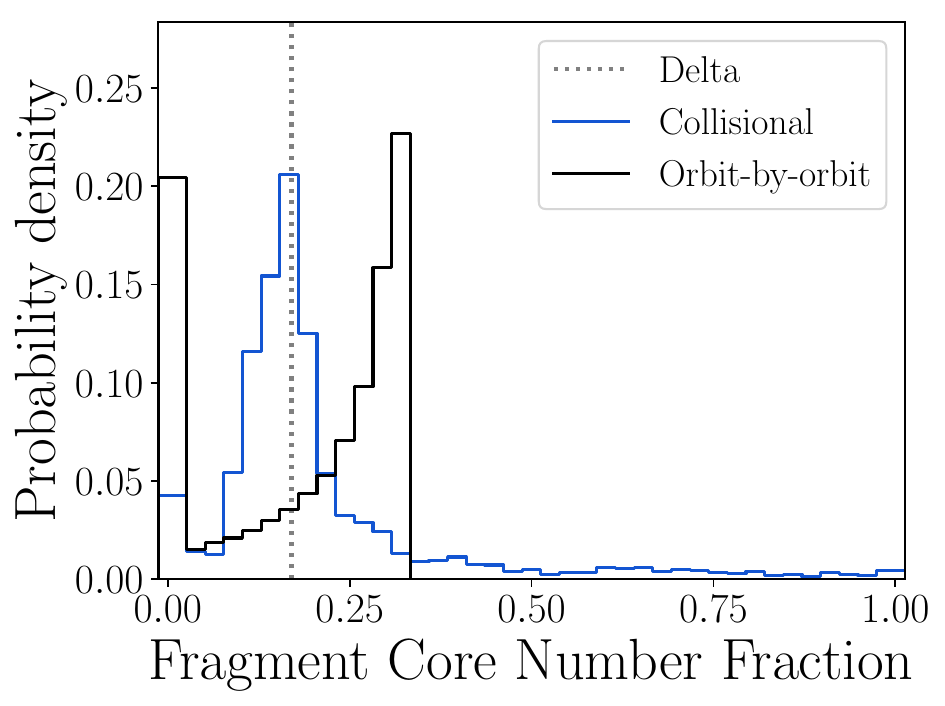}
    \caption[Illustration of the fragment core number fraction distributions predicted by three accretion models]{Illustration of the three fragment core fraction, $f_{\rm c}$, distributions used in this work. Each distribution corresponds to a different accretion model, which we aim to ultimately disentangle by estimating the $f_{\rm c}$ for a large number of synthetic white dwarfs. For details, see Section~\ref{sec:fcnf_dists}.}
    \label{fig:fcnf_dists}
\end{figure}

We assume that the disrupted rocky asteroid has a core mass fraction of 0.35 (corresponding to $f_{\rm c} \approx 0.18$). The innermost and outermost orbits consist entirely of mantle material, because the parts of the asteroid closest to and furthest from the white dwarf at the point of disruption are pure mantle. Intermediate orbits host both core and mantle material. It is assumed that no material is ejected from the system, which is true for sufficiently small asteroids ($\le100$~km for a \SI{10}{AU} semi-major axis orbit, \citealt{Brouwers2023a}). Material is accreted orbit-by-orbit. leading to asynchronous accretion of the core and mantle. The resulting $f_{\rm c}$ distribution is determined by geometry. Consider dividing the disrupted asteroid into slices perpendicular to the axis joining the white dwarf and the asteroid. Let $x$ be the distance along this axis, with the origin at the centre of the asteroid. The volume of each slice is given by

\begin{equation}
    V = \pi\left((x_2-x_1)R^2-\frac{x^3_2-x^3_1}{3}\right) ,
\label{eq:volume_slice}
\end{equation} where $V$ is volume, $R$ is the asteroid radius, and the slice is bounded by $x_1$ and $x_2$. A similar calculation is performed to find the volume of core in each slice. The ratio of these volumes corresponds to one $f_{\rm c}$ value, which is weighted (by volume) and binned to generate the distribution in Figure~\ref{fig:fcnf_dists}. The pure mantle material in the innermost and outermost orbits causes a spike at $f_{\rm c} = 0$, and the maximum $f_{\rm c}$ is given by the core fraction in a slice at $x=0$.

\subsubsection{Delta}

The `Delta' distribution represents a control case, in which no core-mantle differentiation has taken place, and consequently all accreted fragments have the same $f_{\rm c}$ of 0.17. This is an approximately Earth-like value, which we assume is a good proxy for planetesimal core fraction at the time of formation. The value of $f_{\rm c}$ used here differs slightly from the other cases, but this difference is not significant (see Section~\ref{sec:tidalcnf}). This distribution could also correspond to differentiated material which has experienced minimal collisional (or other) processing since formation. This is observationally indistinguishable from undifferentiated material.

\subsubsection{Distinguishing features}

The exact shapes of the collisional and orbit-by-orbit $f_{\rm c}$ distributions are influenced by a number of assumptions and free parameters (which we discuss further in Section~\ref{sec:tidalcnf}). However, we expect the key distinguishing features to be robust. The first feature is a significant tail of core-rich material in the case of collisional disruptions. The second is the bimodal shape of the orbit-by-orbit distribution, which contrasts with the single peak of the collisional distribution. These features are illustrated in Figure~\ref{fig:fcnf_dists}.

\subsection{Observing the synthetic population}
\label{sec:synthetic_observation}

After generating a synthetic white dwarf, and using our forward model to calculate its "true" atmospheric metal abundances, the next step is to mimic the \textit{detected} abundances. Detected elemental abundances inevitably differ from true abundances due to multiple factors. Spectroscopic observations of white dwarfs from \textit{HST}, SDSS or other facilities are used to identify absorption features from metals in the white dwarf atmosphere. The presence and strength of these features depends on the wavelength range used, with narrower and smaller features easier to identify in higher resolution spectra. However, not all elements present will necessarily be identified, typically due to a lack of strong lines in the wavelength range used, or because lines are hidden below the detection threshold.

The abundance of an element is quantified using the equivalent widths (EW) of its absorption lines via atmospheric modelling. The resulting abundance estimates come with associated uncertainties. We assume this uncertainty is purely statistical, and so we mimic it by adding random noise to the synthetic elemental abundances. We determine the detected abundance by randomly sampling from a normal distribution whose mean is the true abundance and whose standard deviation is a free parameter. We assume errors are Gaussian, symmetrical and the same for each element. This is a simplification: errors depend on many factors, such as the number of lines which are detected and the accuracy of line lists, which in general will differ for each element. We also assume the errors on each element are independent (i.e., uncorrelated). In reality, correlations are likely introduced during atmospheric modelling. We discuss these assumptions further in Section~\ref{sec:errors}.

We also mimic the non-detection of elements with low abundances. For each element, we determine the minimum detectable abundance as a function of white dwarf temperature (see section \ref{sec:detectionthresholds} for details). Elements remain undetected if their abundance is below this detection threshold.

\subsubsection{Detection Thresholds}
\label{sec:detectionthresholds}

Our detection thresholds for H-dominated white dwarfs, illustrated in Figure~\ref{fig:detection_thresholds}, were calculated by approximating the minimum detectable abundance as a function of temperature for each element. Firstly, a high resolution ($R\sim40,000$) spectrum was simulated at a signal-to-noise ratio of 30. This corresponds to telescopes and instruments which are commonly used to study polluted white dwarfs, such as KECK/HIRES, Magellan/MIKE and VLT/UVES. By artificially inserting spectral lines, the minimum equivalent width (EW) of the Ca \textsc{ii} K line at 3934\;\AA~ that could be detected to $>3\sigma$ confidence was estimated to be 14\;m\AA~. This EW was converted to a Ca/H abundance ratio using data from from Table 1 in \citet{Zuckerman2003}. This table lists the equivalent width of the Ca \textsc{ii} K line in a sample of DA white dwarfs, along with the inferred Ca abundance and the effective temperature. This process was repeated for the five DAs in \citet{Rogers2024} by scaling the abundances assuming we are in the linear curve of growth regime. We calculate the Ca abundance that corresponds to 14\;m\AA~ EW as a function of effective temperature, and fit a line of best fit to these points. This line is adopted as the detection threshold for Ca in DAs, and is shown in the top left panel of Figure~\ref{fig:detection_thresholds}. It corresponds to a model in which high resolution spectrographs are used, and a signal-to-noise ratio of 30 is targeted.

The detection thresholds for elements other than Ca were calculated by offsetting the abundance by an amount that maintains an EW of 14\;m\AA~ for a typical H-dominated white dwarf in a sample taken from \citet{Rogers2024} (\mbox{Gaia\,J0006+2858}). In principle, these offsets might be affected by temperature. In practise the constant offset appears to match data from real DAs reasonably well, as shown in Figure~\ref{fig:detection_thresholds}.

Our detection thresholds for He-dominated white dwarfs, illustrated in Figure~\ref{fig:detection_thresholds_dz}, were calibrated by eye to match the real detection thresholds in the sample of cool DZs from \citet{Hollands2017}.

Throughout this work, we allow the random noise to vary while detection thresholds remain fixed. In principle, however, the detection thresholds should be linked with the size of the random errors, as both of these are affected by instrument resolution. Higher resolution tends to lead to smaller errors, and less restrictive detection thresholds.

\subsection{Parameter Retrieval}
\label{sec:synthetic_modelling}

The final stage of the pipeline is to consider how the detected elemental abundances would be modelled in terms of the formation history of the pollutant, and compare this against the original parameters used to describe the accreted planetesimals. The property we are most interested in is $f_{\rm c}$, but we must also consider the initial stellar composition and the phase of accretion in order to estimate $f_{\rm c}$. Our approach reuses the key calculations of \citet{Harrison2018}, \citet{Harrison2021} and \citet{Buchan2021}, but simplified and approximated for computational feasibility. We do not perform parameter retrieval on He-dominated white dwarfs due to their additional complexity (see Section~\ref{sec:phaseofaccretion}).

\subsubsection{Phase of accretion}
\label{sec:phaseofaccretion}

In any interpretation of white dwarf pollution, it is crucial to take the (unknown) accretion phase of the system into account because this changes the inferred abundance of every element in the accreted material. We can reasonably assume that H-dominated white dwarfs are in steady state (due to their typically short sinking timescales). In this case, we correct for differential sinking by weighting each abundance by the inverse of its sinking timescale, and proceed using these corrected abundances. For He-dominated white dwarfs, steady state cannot be safely assumed, and we cannot determine the accretion phase because it is degenerate with the initial nebular composition. To disentangle these variables, a more sophisticated, and computationally demanding, model is required (e.g., that of \citealt{Buchan2021} or \citealt{Swan2023a}).

\subsubsection{Determining initial nebular composition}
\label{sec:determing_initial_composition}

The initial composition from which the pollutant formed is assumed to match one of 958 real stars, as described in Section~\ref{sec:initialcomp}. We fit the initial nebular composition by determining which of these 958 real stars best matches the synthetic detected metal abundances. We adopt the composition of that star as the initial nebular composition.

We determine the degree of (mis)match between a real star and a set of synthetic abundances by comparing the relative abundances of Al, Ti, Ca and Mg. These elements are chosen because, in our model, the only process which significantly affects their relative abundances is differential sinking. We should therefore be able to take the abundances of these elements from a synthetic detected system, scale them according to their sinking timescales to undo the effects of differential sinking, and recover their abundances in the initial nebula. In principle, these recovered abundances should exactly match one of the 958 real stars. For each of the 958 real stars, we calculate the mismatch, $M$, as

\begin{equation} 
M = \sum_{X = \textrm{Ti}, \textrm{Ca}, \textrm{Mg}} \max\left(R(X), \frac{1}{R(X)}\right) - 1,
\end{equation} in which $\max(a, b)$ is the largest value out of $a$ and $b$, and
\begin{equation} 
 R(X) = \frac{X_{s}Al_{*}\tau_{Al}}{X_{*}Al_{s}\tau_{X}},
\end{equation} where $X_s$ is the abundance of metal X in the synthetic system, $X_*$ is the abundance of X in the real star, $\tau_{X}$ is the sinking timescale of X in the synthetic system, and Al is chosen as an arbitrary baseline. If the synthetic abundances match the real abundances perfectly, then $R=1$ for all elements and the mismatch $M$ is zero. We identify the star which minimises $M$, and if there are multiple stars which share the joint lowest mismatch, we propagate all their compositions through the retrieval and return multiple values for $f_{\rm c}$, which we assume to be equally likely.

\subsubsection{Determining fragment core fraction}
\label{sec:determing_fcf}

We scale the detected Ca/Fe ratio to compensate for the initial nebular composition determined in Section~\ref{sec:determing_initial_composition}. We use this adjusted ratio to analytically calculate $f_{\rm c}$, assuming Earth-like core and mantle compositions.

If Ca and/or Fe are not present, we use the following ratios instead, in priority order: Mg/Fe, Al/Fe, Ti/Fe, Ca/Ni, Mg/Ni, Al/Ni, Ti/Ni, Ca/Cr, Mg/Cr, Al/Cr, Ti/Cr.

\subsection{Assessing the consistency of different distributions}
\label{sec:ks_testing}

For a sample of synthetic H-dominated systems, the key output from the parameter retrieval stage of the pipeline is a $f_{\rm c}$ distribution across the whole sample. In order to determine whether this distribution is statistically distinguishable from another distribution (e.g., a real $f_{\rm c}$ distribution), we perform a two-sample Kolmogorov–Smirnov (KS) test. The KS test proceeds by calculating

\begin{equation}
    D = \max(|C_1 - C_2|),
\end{equation}
where $C_1$ and $C_2$ are the $f_{\rm c}$ distributions expressed as cumulative distributions, and $D$ can be visualised as the maximum vertical distance between them. An associated p-value can also be calculated to test the null hypothesis that the two samples are drawn from the same underlying distribution. If the p-value is lower than some threshold $\alpha$, the null hypothesis is rejected and the samples are deemed to be distinguishable. We set $\alpha=0.05$. Whilst this threshold is commonly used, it is ultimately arbitrary. We perform this test multiple times to calculate the probability that two populations can be distinguished by drawing samples from them. We call this probability $P(X \neq Y)$, where X and Y are the underlying distributions.

For synthetic He-dominated samples, we do not perform parameter retrieval, and so the key outputs are the distributions of detected abundances for key elements. We use a Cramér test, with a \texttt{phiCramer} kernel \citep{Baringhaus2004}, to quantify the consistency of these distributions with real equivalents. This test functions similarly to a KS test, but is multivariate and therefore suitable for simultaneous comparison of (relative) Ca, Mg and Fe abundances.

\subsection{Model verification}
\label{sec:control}

\begin{figure}
    \centering
    \includegraphics[width=\columnwidth,keepaspectratio=true]{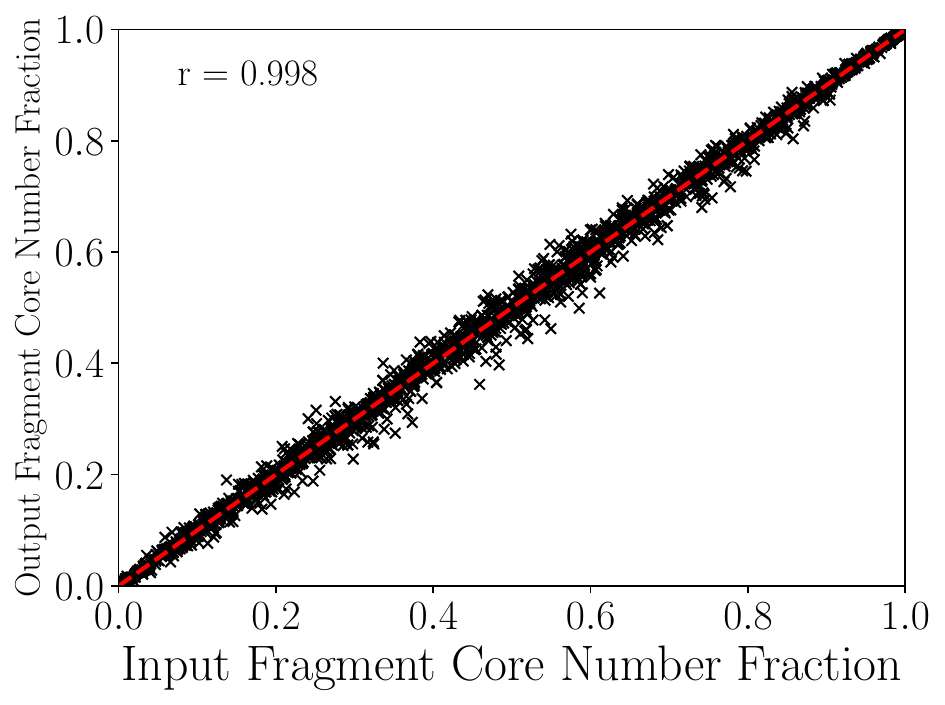}
    \caption[Validation of the parameter retrieval algorithm as applied to population of synthetic white dwarfs]{A test of how well parameter retrieval performs under control conditions (i.e., no random noise, bias or missing elements). We set up a control population of white dwarfs as described in Section~\ref{sec:control}. The white dwarfs are polluted with material of varying fragment core fraction, $f_{\rm c}$, shown on the x axis. The y axis shows the corresponding $f_{\rm c}$ retrieved by our algorithm. The strong correlation demonstrates that the parameter retrieval component of the pipeline works well. Retrieval failed for 2 of the 2000 systems with an extremely low $f_{\rm c}$. The red dashed line illustrates a perfect 1:1 correlation.}
    \label{fig:ControlIOScatter}
\end{figure}

The parameter retrieval model was tested using a synthetic population of 2000 H-dominated white dwarfs. The input values are listed in Table~\ref{tab:synthetic_parameters}. For testing purposes only, the $f_{\rm c}$ distribution used was uniform from 0 to 1. The model retrieves $f_{\rm c}$ for each synthetic white dwarf based on its (mock) detected metal abundances, assuming no random noise or missing elements. It achieves a high degree of accuracy, as illustrated by Figure~\ref{fig:ControlIOScatter}. The output values of $f_{\rm c}$ are strongly correlated with the input values, across the full range of possible input values, with a Pearson correlation coefficient of 0.998. 

The model failed to retrieve the core number fraction in 2/2000 systems with extremely low $f_{\rm c}$, and additionally that the model is not always able to retrieve the initial nebular composition well. Neither of these deficiencies has much effect on our results. The initial nebular composition is only used as a stepping stone to estimate $f_{\rm c}$, which is generally retrieved accurately.

\section{Results}
\label{sec:synthetic_results}

This work presents forward models that predict the distribution of compositions in a sample of polluted white dwarfs, depending on three possible accretion scenarios (denoted Orbit-by-orbit, Collisional and Delta, as explained in Section~\ref{sec:fcnf_dists}). Here we firstly consider a representative example that illustrates the effects of detection thresholds and random noise (Sections~\ref{sec:controlrealistic} and \ref{sec:controlrealisticnoise}). We then consider samples of white dwarfs generated according to the three different accretion models, investigating the conditions under which they may be distinguished (Section~\ref{sec:fcf_comparison}). Finally, we consider whether any of these models is favoured by a sample of cool DZs (Section~\ref{sec:hollands}). The setup of each section is summarised in Table~\ref{tab:results_desc}.

\begin{table*}
\centering
\caption[Table summarising different pipeline set-ups]{Summary of the pipeline set-up for the various results sections. $b$ refers to the parameter in Equation~\ref{eq:pol_frac}.}
\label{tab:results_desc}
\begin{tabular}{c c c c c c}
\hline
Section & $f_{\rm c}$ Distribution(s) & WD Parameters & Detection Thresholds & Noise & $b$\\ 
\hline
\ref{sec:control}, \ref{sec:controlrealistic} & Uniform & 40pc DAs & \checkmark & - & $-3$ \\
\ref{sec:controlrealisticnoise} & Uniform & 40pc DAs & \checkmark & \checkmark & $-3$ \\
\ref{sec:fcf_comparison} & Orbit-by-orbit, Collisional, Delta & 40pc DAs & \checkmark & \checkmark & $-3$ \\
\ref{sec:hollands} & Orbit-by-orbit, Collisional, Delta & \citet{Hollands2017} & \checkmark & \checkmark & $-5$ \\
\end{tabular}
\end{table*}

\subsection{The effect of non-detection of elements}
\label{sec:controlrealistic}

\begin{figure}
    \centering
    \includegraphics[width=\columnwidth,keepaspectratio=true]{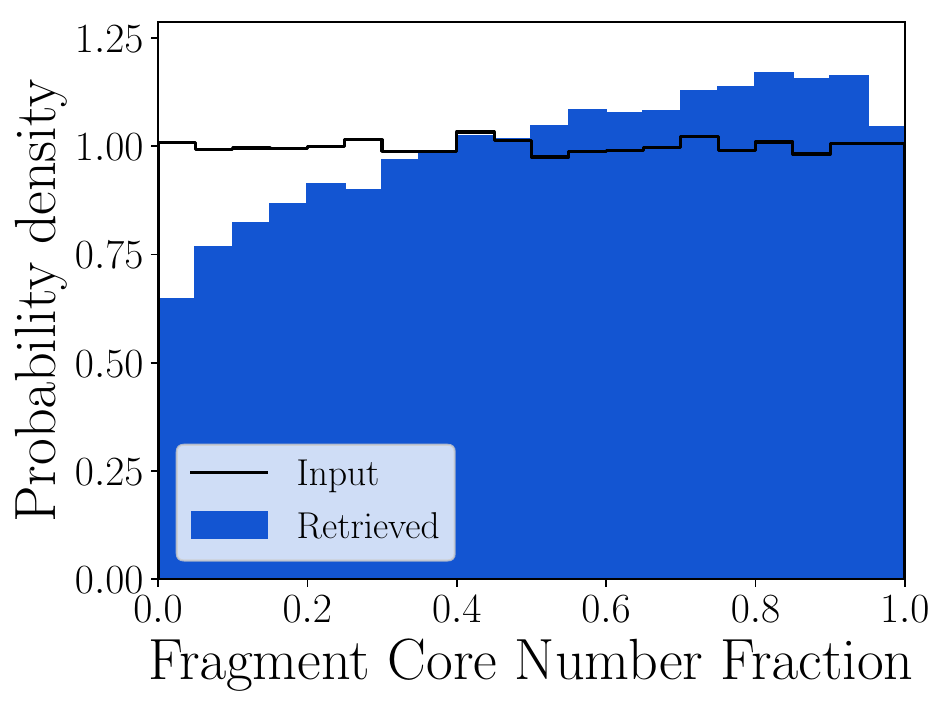}
    \caption[Example of a synthetic population of white dwarfs biased towards core-rich detections]{An example of how the deduction of core content in planetesimals accreted on to white dwarfs can be biased, based on mock detections of elements in white dwarf photospheres. The blue bars show the retrieved distribution of fragment core fraction, $f_{\rm c}$, for material accreted onto a synthetic white dwarf population when using realistic detection thresholds, but no random noise. Abundances of Ca, Fe (and other backup elements) below these thresholds were considered to not be detectable (see Section~\ref{sec:detectionthresholds}). The output population is skewed towards core-rich systems due to the resulting preferential dropout of core-poor systems relative to the initial population (black line). Extremely core-rich systems are also preferentially removed to a lesser extent. The peak around 0.8 is a result of subsequent normalisation. Retrieval was possible for 34,432/100,000 systems.}
    \label{fig:ControlIOFcfDistRealistic}
\end{figure}

\begin{figure}
    \centering    \includegraphics[width=\columnwidth,keepaspectratio=true]{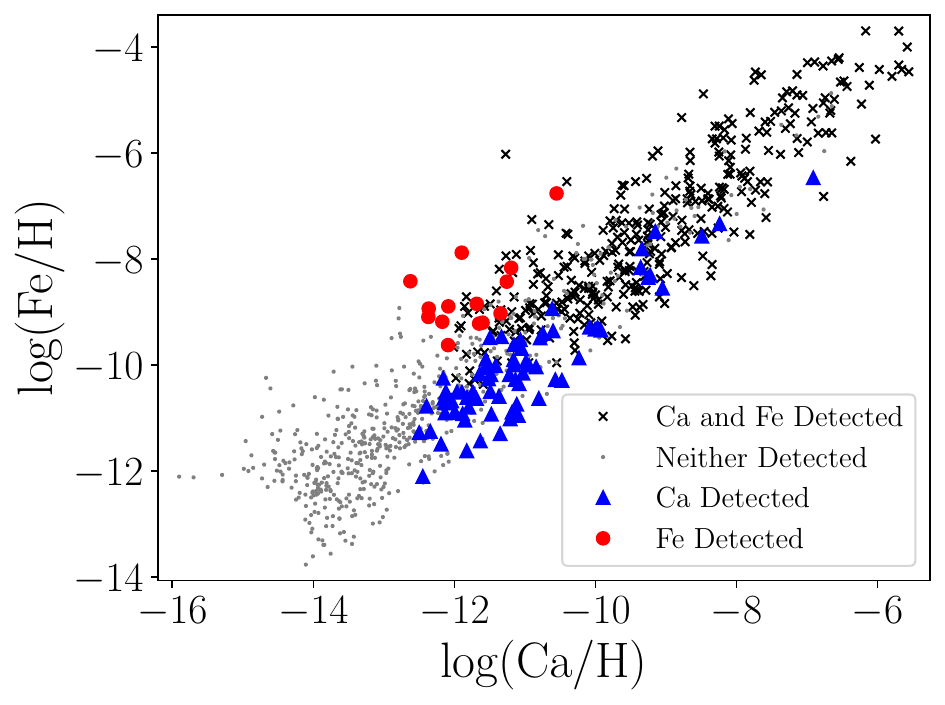}
    \caption[Detected Ca and Fe abundances for 1000 synthetic white dwarfs]{Detected Ca and Fe abundances in the photospheres of 1000 synthetic white dwarfs selected randomly from the population described in Section~\ref{sec:controlrealistic}. To detect both of these elements, their abundances must exceed a certain threshold determined by the individual white dwarf's temperature (see Section~\ref{sec:detectionthresholds}). In this case, a fragment core number fraction can be estimated (black crosses), and the system contributes to the fragment core fraction, $f_{\rm c}$, distribution. Systems can drop out of the $f_{\rm c}$ distribution if neither Ca or Fe is detected (black circles), or if only one of Ca (blue crosses) and Fe (red crosses) is detected (unless backup elements are detected). These systems introduce bias because it is more likely that mantle-rich systems, with high Ca/Fe, drop out of the $f_{\rm c}$ distribution than the core-rich systems (i.e., the blue crosses outnumber the red crosses).}
    \label{fig:CaFeScatterControlRealistic}
\end{figure}

It is almost always harder to detect Fe than Ca or Mg in the optical. In this case, the detected sample is biased towards more (apparently) core-rich pollutants. This is illustrated in Figure~\ref{fig:ControlIOFcfDistRealistic}, which shows the retrieved $f_{\rm c}$ distribution for a population of 100,000 H-dominated white dwarfs. This population was generated as in Section~\ref{sec:control}, with a uniform $f_{\rm c}$ distribution (black line in Figure~\ref{fig:ControlIOFcfDistRealistic}). The core-rich bias is due to the preferential dropout of mantle-rich systems, illustrated in Figure~\ref{fig:CaFeScatterControlRealistic} (using Ca and Fe as proxies for mantle and core). The red and blue crosses show systems where only one of Ca and Fe is detected. For these systems, $f_{\rm c}$ cannot be inferred and so they drop out of the blue distribution in Figure~\ref{fig:ControlIOFcfDistRealistic} (unless backup elements happen to be detected). Systems with Ca detected and Fe undetected are more common than the inverse and have relatively high Ca/Fe (i.e., are more mantle-rich). The 34,432 systems for which a $f_{\rm c}$ can be retrieved are modelled accurately (r = 0.995).

\subsection{The effect of random noise }
\label{sec:controlrealisticnoise}

We now repeat the test in Section~\ref{sec:controlrealistic}, adding random Gaussian noise.

\begin{figure}
    \centering
    \includegraphics[width=\columnwidth,keepaspectratio=true]{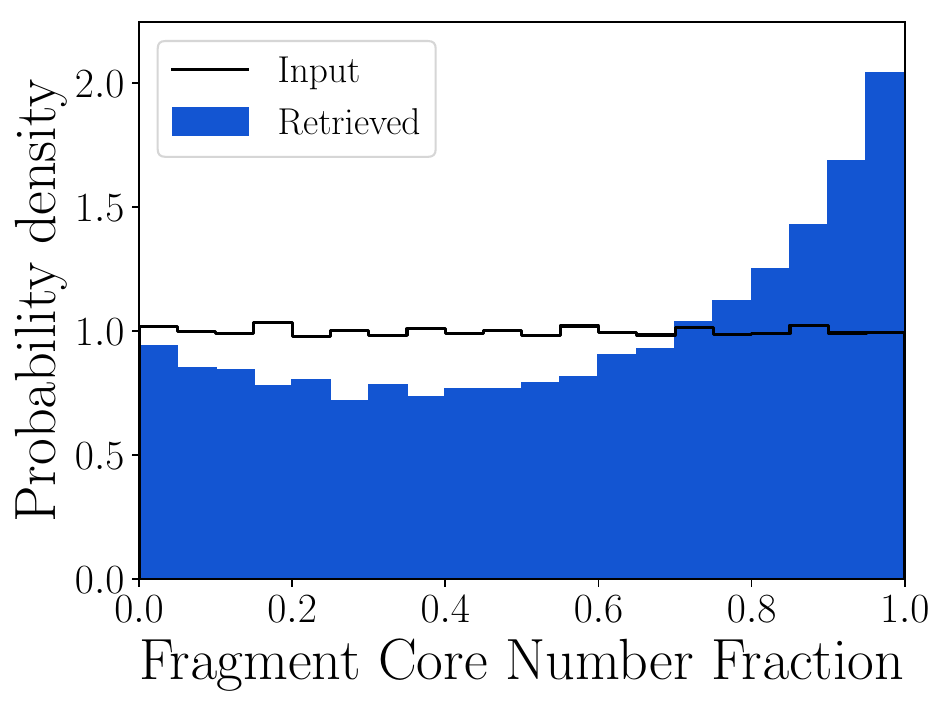}
    \caption[Example of a synthetic population of white dwarfs biased towards detection of extreme compositions by the addition of random noise]{Similar to Figure~\ref{fig:ControlIOFcfDistRealistic}, but with the addition of 0.4\;dex errors due to random noise. The resulting detected $f_{\rm c}$ distribution (blue bars) is pushed towards extreme values (i.e., mantle-rich and core-rich objects) relative to both the input distribution (black line) and the noise-free equivalent (Figure~\ref{fig:ControlIOFcfDistRealistic}). Retrieval was possible for 34,031/100,000 systems (primarily limited by whether pollution levels are high enough to be detectable).}
    \label{fig:ControlNoiseIllustration}
\end{figure}

The effect of adding noise is, in general, to push the inferred fragment core fraction towards extreme values. As an example, Figure~\ref{fig:ControlNoiseIllustration} shows how an initially uniform $f_{\rm c}$ distribution is distorted towards extreme values when the Gaussian noise is 0.4 dex. This occurs because $f_{\rm c}$ is inferred from a ratio of two elemental abundances (or rather, the difference, since the errors are applied in log space). As the random error becomes very large, this difference is dominated by the error, and will become either very positive or very negative. When transformed back into linear space, the ratio tends to become either very small or very large, resulting in a fragment core number fraction close to either 0 or 1.

Figure~\ref{fig:ControlNoiseIllustration} includes the effects of both detection thresholds (which remove extreme values) and noise (which artificially replaces them). The effects do not exactly cancel out at any point, however. The detection thresholds are more effective at removing low fragment core fractions than high ones, while random noise artificially enhances both ends of the distribution.

\subsection{Estimating the sample size needed to distinguish accretion models}
\label{sec:fcf_comparison}

We now propagate three different populations of H-dominated white dwarfs through the pipeline, corresponding to each of the three accretion models (Orbit-by-orbit, Collisional, Delta). These models are defined, for our purposes, by fragment core fraction distributions (see Section~\ref{sec:fcnf_dists}). We include detection thresholds and vary the level of noise.

\begin{figure*}
    \centering
    \includegraphics[width=\textwidth,keepaspectratio=true]{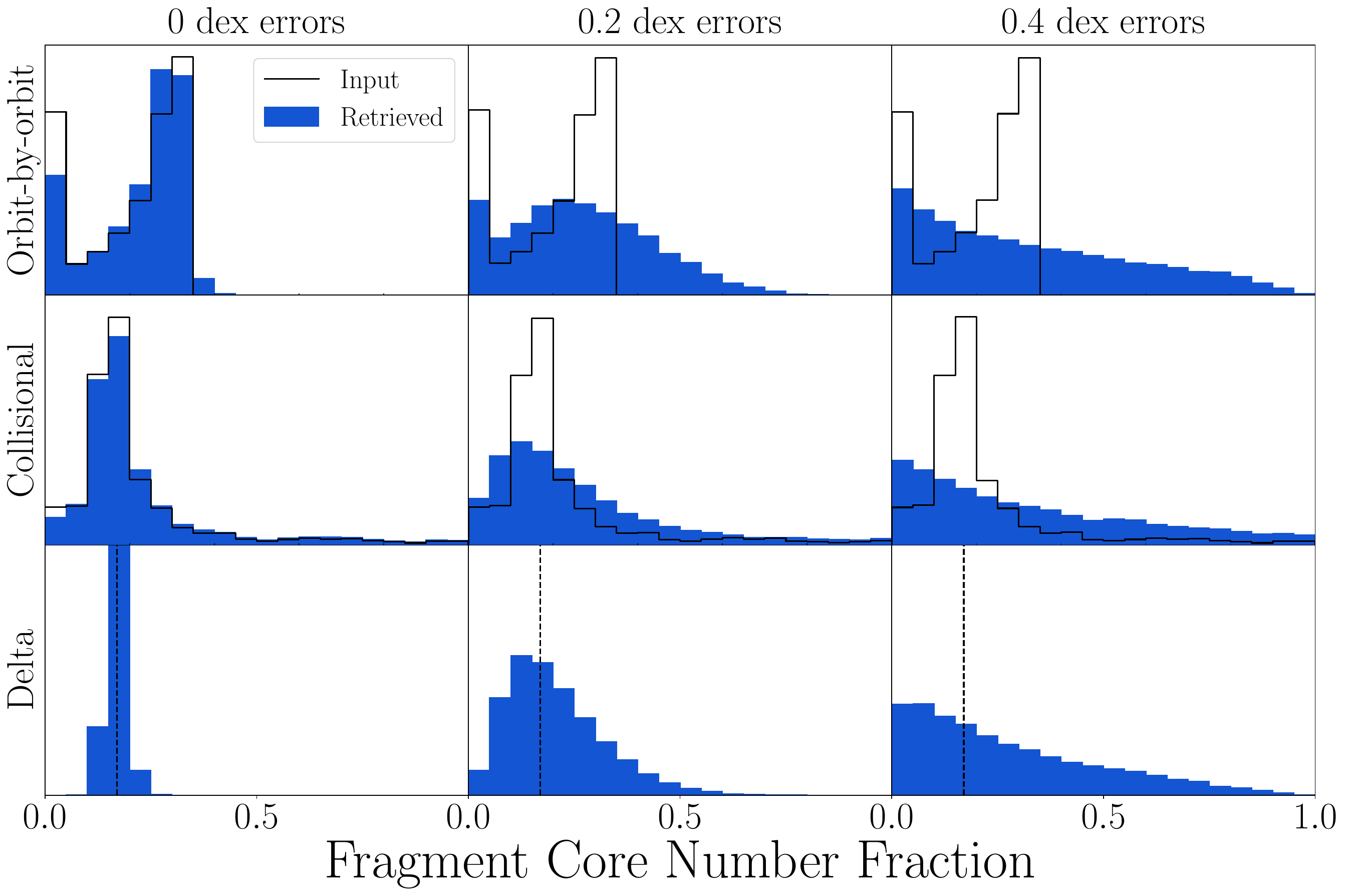}
    \caption[Retrieved fragment core fraction distributions for three different accretion models at three different noise levels]{The effect of adding random noise is to broaden, and eventually wash out, the true distribution of the composition of accreted material. Here, composition is defined by the fragment core number fraction: the fraction of accreted material which is core-like. From left to right, each panel in a row corresponds to increasing the random noise injected into a synthetic population of white dwarfs. The true fragment core fraction, $f_{\rm c}$, distribution is indicated with a black line, and the retrieved distribution with blue bars. The three rows correspond to the Orbit-by-orbit, Collisional and Delta $f_{\rm c}$ distributions described in Section~\ref{sec:fcnf_dists}. When no random noise is added (left column), these distributions are retrieved accurately. When 0.2\;dex errors are added (middle column), none of the retrieved distributions closely resemble the corresponding input, but can still be distinguished by the number of peaks and the range of values covered. When the errors are increased to 0.4 dex, these distinguishing features are largely washed out. The input distributions show all 100,000 generated systems in the relevant population, with the exception of the `Delta' case for which the single input value is indicated with a vertical line. The retrieved values are only shown for the systems for which parameter retrieval was possible (just under 30,000 systems in each case).}
    \label{fig:tidalvcollisional}
\end{figure*}

We illustrate the retrieved distributions in Figure~\ref{fig:tidalvcollisional}. The rows correspond to the different accretion models, defined by the initial $f_{\rm c}$ distribution (black line). The blue bars illustrate the retrieved (final) distribution. The columns correspond to different noise levels. The left column shows the case of zero random noise, demonstrating that the model recovers the initial distributions well in this case (blue bars). The only notable discrepancy is a bias against retrieval of very low $f_{\rm c}$ (this is due to detection thresholds; see Section~\ref{sec:controlrealistic}).

Adding noise (middle and right hand columns) causes the retrieved distributions to become smeared out. When the errors are 0.2 dex (middle column) the populations can still be visually distinguished by the number of peaks and presence or absence of an extended core-rich tail. This is true despite the fact that no distribution is retrieved to a high degree of accuracy. However, these features are largely washed out once the noise is increased to 0.4 dex (right column).

\begin{figure}
    \centering
    \includegraphics[width=0.9\columnwidth,keepaspectratio=true]{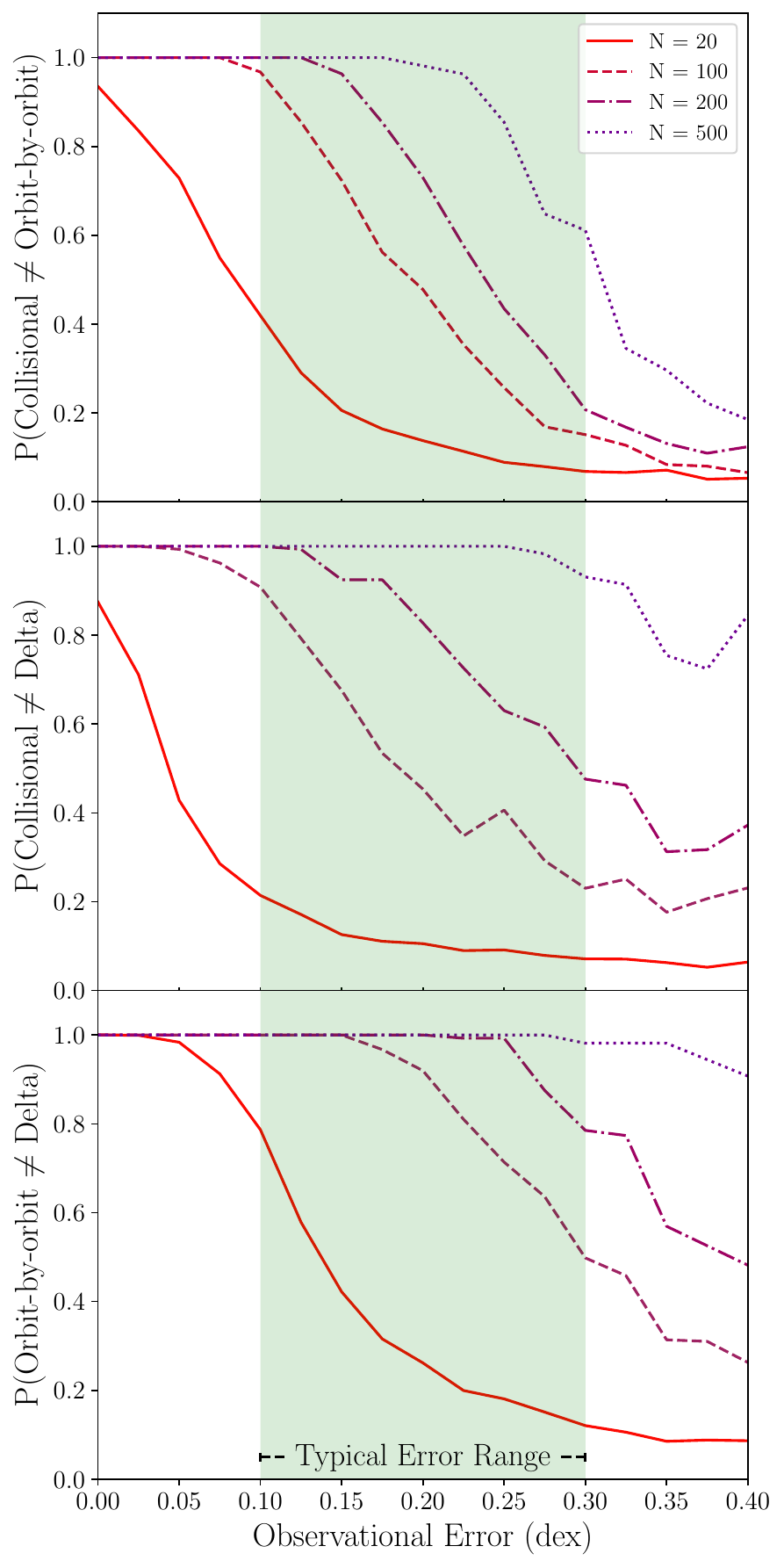}
    \caption[Probability that accretion models can be statistically distinguished as a function of sample size and error due to random noise]{Top: Probability that two samples of accreted pollutants, following Collisional and Orbit-by-orbit accretion models, can be distinguished statistically as a function of sample size and error due to random noise. Middle and Bottom: Similar to top panel, but comparing the Delta model to Collisional (middle) or Orbit-by-orbit (bottom). Given realistic errors of 0.2 dex, a sample size on the order of 500 allows all three models to be reliably distinguished from each other. The green shaded region shows the typical 1 sigma error range based on Ca detections for all systems for which we have data. Note that these sample sizes assume that each white dwarf in the sample has a $f_{\rm c}$ estimate (which requires detection of multiple elements). In our synthetic populations, this was possible for roughly half of the white dwarfs with detectable pollution, or about 30\% of the whole population.}
    \label{fig:fcf_comparisons}
\end{figure}

The key question is: how noisy can the data be before these distributions can no longer be distinguished, and what is the effect of sample size? For a given noise level and sample size, we calculate the probability that these distributions can be distinguished, $P(\textrm{Distribution 1} \neq \textrm{Distribution 2})$, via KS testing (see Section~\ref{sec:ks_testing}). We illustrate the dependence of this probability on noise and sample size in Figure~\ref{fig:fcf_comparisons} for each combination of distributions. We draw independent samples from a pool of 100,000 until we run out of systems. Systems were ignored if $f_{\rm c}$ could not be inferred: these sample sizes refer to a sample of white dwarfs each with individual estimates of $f_{\rm c}$, which requires detection of multiple elements.

When there is no noise, any two distributions can usually be distinguished even for a small sample size of 20. As the noise level increases while holding sample size fixed at 20, it generally becomes harder to distinguish them. Given large enough observational error, it is not possible to confidently distinguish any of the underlying distributions from each other. This effect can be compensated for by increasing sample size.

The top panel tracks $P(\textrm{Collisional} \neq \textrm{Orbit-by-orbit})$, which drops below 50\% for a sample size of 100 once the observational error is increased to a realistic value of 0.2 dex. Given this noise level, we find that a sample size of 500 is sufficient to ensure distinguishability (98\%). To reach a 90\% chance of distinguishability, we estimate that a sample size of 275 is necessary.

The middle and bottom panels of Figure~\ref{fig:fcf_comparisons} track $P(\textrm{Collisional} \neq \textrm{Delta})$ and $P(\textrm{Orbit-by-orbit} \neq \textrm{Delta})$. Both these cases test whether differentiated material can be distinguished from undifferentiated material. Given 0.2 dex errors, a sample size of 500 ensured that the Delta model could be distinguished from Collisional in every test, while the equivalent figure for the Orbit-by-orbit model is 200. We estimate that sample sizes of 275 (Collisional) and 100 (Orbit-by-orbit) are sufficient for a 90\% chance of distinguishability from the Delta model.

\begin{figure}
    \centering
    \includegraphics[width=\columnwidth,keepaspectratio=true]{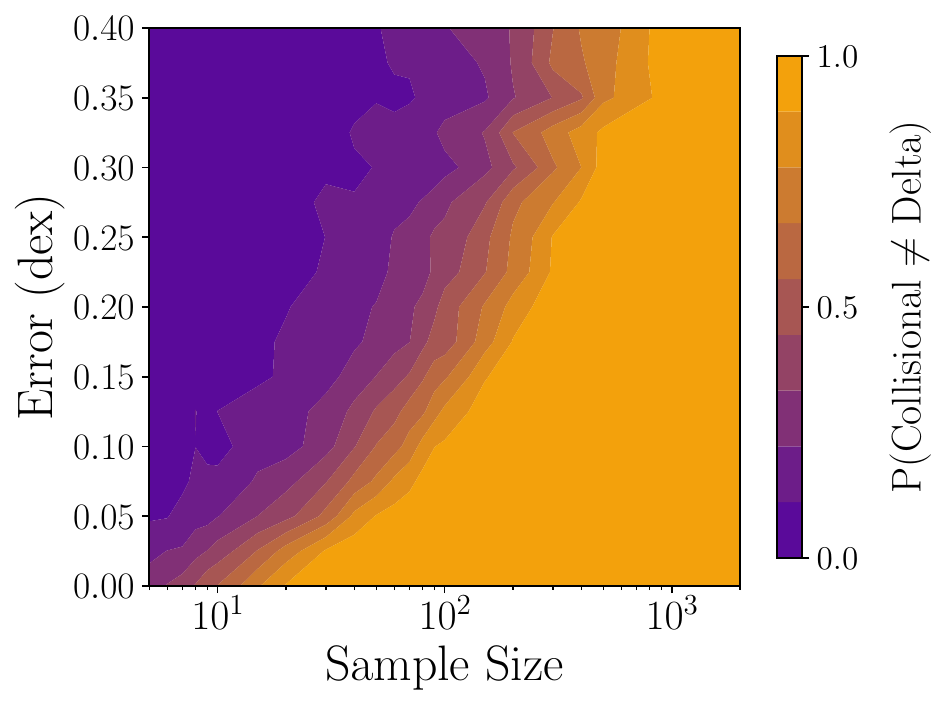}
    \caption{The probability that two samples of polluted white dwarfs can be distinguished as a function of error (i.e., random noise) and sample size. One sample corresponds to collisionally evolved pollutants, the other to pollutants which follow a Delta $f_{\rm c}$ distribution. This plot is essentially a different representation of the middle panel of Figure~\ref{fig:fcf_comparisons}. As in Figure~\ref{fig:fcf_comparisons}, the sample size includes only systems for which $f_{\rm c}$ can be estimated, which requires detection of multiple elements.}
    \label{fig:dprobheatmap}
\end{figure}

Figure~\ref{fig:dprobheatmap} shows how $P(\textrm{Collisional} \neq \textrm{Delta})$ varies as a function of both error and sample size. This is illustrative of the trade-off between error and sample size: for example, a sample of 60 white dwarfs with a typical error of 0.1 dex has similar statistical power to a sample of 200 white dwarfs with 0.3 dex errors.

These estimates do not explicitly consider instrument resolution because this is tied into the noise level (see Section~\ref{sec:future_surveys}).

\subsection{Application to a sample of cool DZs}
\label{sec:hollands}

Figure~\ref{fig:fcf_comparisons} suggests that it is necessary to analyse populations of $\gtrsim$100 polluted white dwarfs (with detections of multiple relevant elements) for evidence that one of the three accretion models considered here is favoured over the other two. We next turn our attention to a sample of 202 cool white dwarfs with He-dominated atmospheres. The sample is taken from \citet{Hollands2017}, with updated Mg abundances from \citet{Blouin2020}. The sample is not necessarily selected in a manner that is unbiased in terms of composition, so caution should be taken in analysing the distribution of compositions. Each white dwarf in this sample has detections of at least Ca, Mg and Fe. \citet{Harrison2021} found that 64 of these systems show evidence of core-mantle differentiation, 7 of which are to $>3 \sigma$. However, these results could conceivably be attributed to random noise (i.e., the Delta model in Section~\ref{sec:fcnf_dists}). We investigate whether this white dwarf sample is consistent with a Delta distribution. We also test for consistency with the Orbit-by-orbit and Collisional distributions described in Section~\ref{sec:fcnf_dists}.

This analysis differs from Section~\ref{sec:fcf_comparison} in two key ways. Firstly, we are testing synthetic samples for consistency with a real sample (of fixed sample size), rather than testing them against each other. Secondly, this sample contains He-dominated (rather than H-dominated) white dwarfs. This introduces additional uncertainty from differential sinking, and parameter retrieval cannot be carried out (see Section~\ref{sec:synthetic_modelling}). We instead compare the relative abundances of Ca, Fe and Mg between the synthetic and real samples. The different evolutionary histories should manifest as different distributions of the abundances of these elements. Population synthesis and mock detection is also modified for He-dominated systems, as detailed in Sections~\ref{sec:synthetic_generation} and \ref{sec:synthetic_observation}.

\begin{figure*}
    \centering
    \includegraphics[width=\textwidth,keepaspectratio=true]{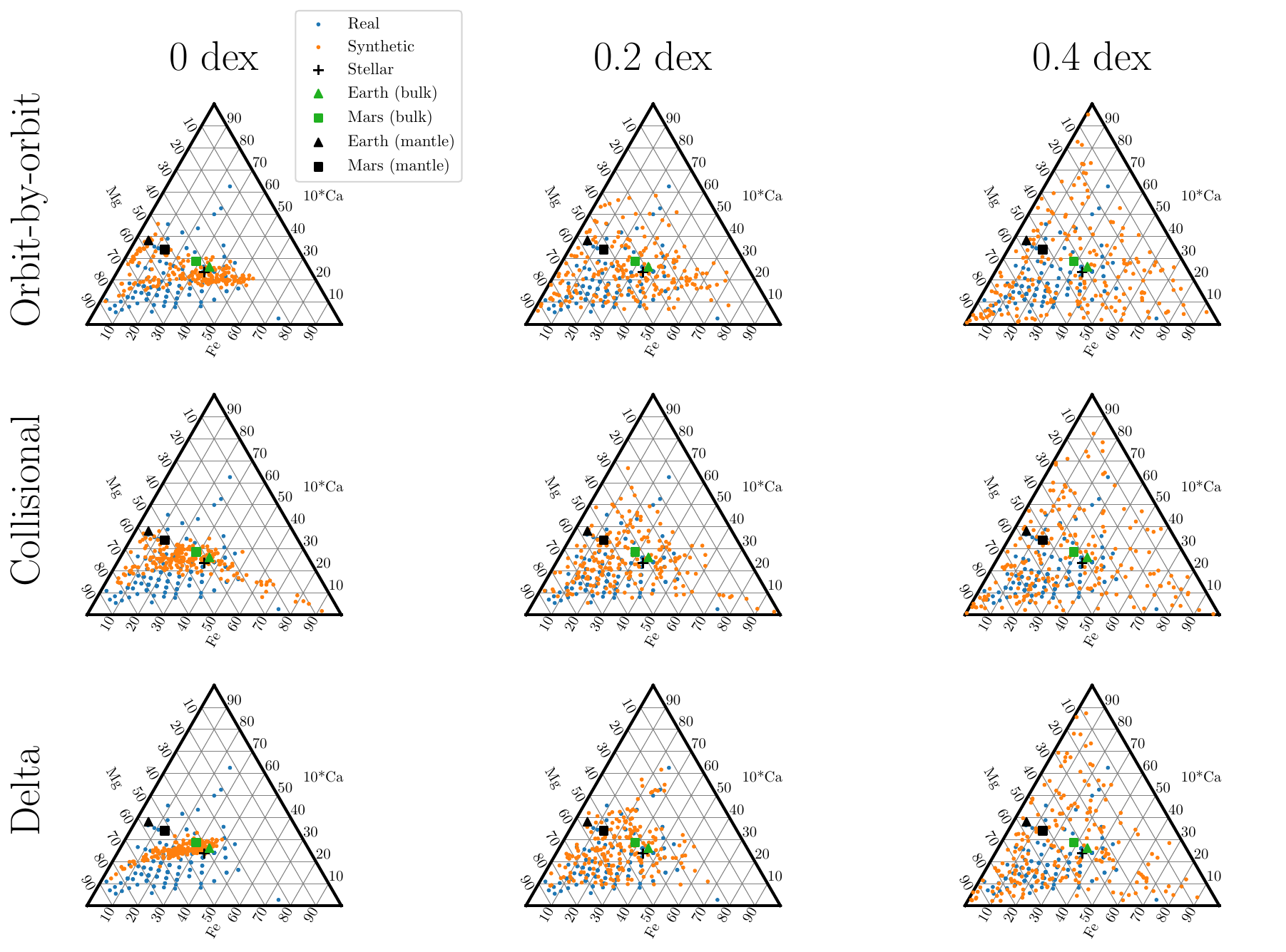}
    \caption{Ternary plots comparing the relative Ca, Mg and Fe abundances of real (blue) and synthetic (orange) white dwarfs. The real abundances are taken from a sample of 202 cool DZs \citep{Hollands2017,Blouin2020}. Some of these systems have the same relative Ca, Mg and Fe abundances. Each of the nine panels shows a different set of synthetic abundances, each with 202 data points. The nine synthetic sets are generated using different combinations of underlying $f_{\rm c}$ distribution (rows) and level of random noise (columns). The synthetic samples are randomly selected from those systems which have detectable levels of Ca, Mg and Fe. The abundance of Ca is multiplied by 10 for visual clarity (for purposes of this figure only). The relative Mg and Fe abundances are therefore higher than shown. By eye, the synthetic data best matches the real data at a noise level of 0.2 dex. For this noise level, the Delta model appears to offer a better match to the data than either of the Collisional or Orbit-by-orbit models. Stellar composition is shown for reference, which is a good proxy for chondritic composition. The bulk and mantle compositions of Earth and Mars are also shown \citep{McDonough2003,Yoshizaki2020}. The core content of Ca and Mg is assumed to be zero.}
    \label{fig:ternaryplots}
\end{figure*}

We create three synthetic populations of 50,000 cool He-dominated white dwarfs, corresponding to each of the $f_{\rm c}$ distributions described in Section~\ref{sec:fcnf_dists}. Figure \ref{fig:ternaryplots} compares the real Ca, Fe and Mg abundances against the synthetic abundances for each of these populations, and for each of three different noise levels (0, 0.2 and 0.4 dex). Each sample (real or synthetic) contains 202 data points, although some real data points occupy the same location due to rounding. The 202 synthetic data points are randomly drawn from the subset of all systems with detectable abundances of Ca, Mg and Fe. Visual inspection suggests that, of the noise levels shown, 0.2 dex errors permit the best match between synthetic and real data. For this noise level, the Delta distribution appears to offer a better match to the data than either of the Collisional or Orbit-by-orbit distributions.

\begin{figure*}
    \centering
    \includegraphics[width=\textwidth,keepaspectratio=true]{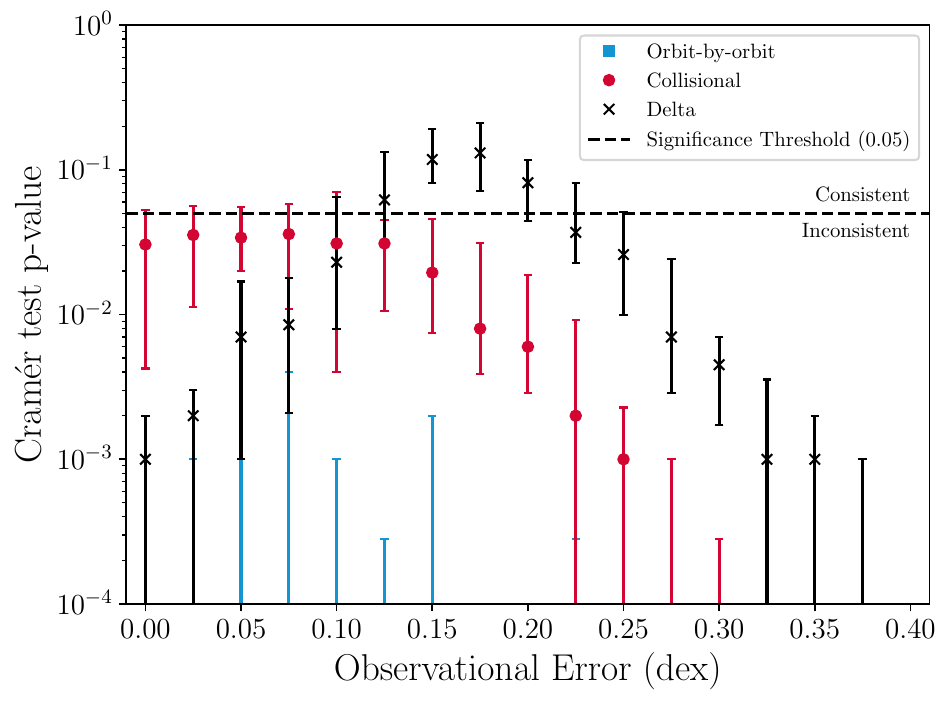}
    \caption{A measurement of how well the synthetic populations of white dwarfs match the sample of cool DZs in \citet{Hollands2017} as a function of random noise. The vertical axis indicates the p-value calculated from a Cramér test as applied to the relative abundances of Ca, Mg and Fe. Higher values indicate greater consistency with the real data. The horizontal dashed line indicates the 0.05 significance threshold, below which the real and synthetic samples are inconsistent. This threshold is arbitrary and should not be treated as a hard boundary. Results for three $f_{\rm c}$ distributions are shown: Orbit-by-orbit (blue squares), Collisional (red circles) and Delta (black crosses). The orbit-by-orbit p-values are extremely low, so are cropped out for visual clarity. Reasonable matches to the data are found for the Delta distribution (with random noise between about 0.1 and 0.25 dex) and for the Collisional distribution with noise levels below about 0.15 dex.}
    \label{fig:SyntheticHollandsCramer}
\end{figure*}

We quantify this result by performing a Cramér test (see Section~\ref{sec:ks_testing}). This test is competitive with other multivariate tests in terms of sensitivity to the dispersion and location of test distributions \citep{Puritz2022}. Importantly, it also proved sensitive to changes in our synthetic compositional distributions in practice. Figure~\ref{fig:SyntheticHollandsCramer} illustrates the p-values obtained, including 1 sigma error bar estimates based on generating as many independent synthetic samples as possible, given the total number of synthetic systems.

On average, the best match to real data is obtained for synthetic populations generated from a Delta distribution, assuming 0.175 dex errors (although reasonable matches can be obtained for noise levels between about 0.1 and 0.25 dex). The Collisional distribution can also yield acceptable matches to the data, although it requires lower noise levels (below $\sim0.15$ dex) as it is an inherently broader distribution.

While the Cramér test generally appears to capture the degree of visual compatibility between real and synthetic data well, we note that for very low noise levels ($<0.1$ dex), the Collisional samples do not appear to match very well visually despite being statistically consistent. However, such noise levels are likely unrealistic in any case.

\section{Discussion}
\label{sec:synthetic_discussion}

We find three key results. First, any realistic sample of polluted white dwarfs is likely biased towards core-rich material (with low Ca/Fe). Secondly, a sample of 500 polluted H-rich white dwarfs (each with multiple elements, such that core and mantle content can be inferred) is large enough to reliably distinguish between Collisional, Orbit-by-orbit and Delta accretion scenarios. Finally, the sample of 202 cool DZs from \citet{Hollands2017} is sufficiently noisy that there is no evidence for a an additional spread in Ca, Fe and Mg abundances beyond the presence of Gaussian noise. Consequently, the data are best explained by our Delta model, in which the accreted material is undifferentiated (or unfragmented). We begin by discussing the implications of this result.

\subsection{The cool DZ sample}

Our results demonstrate that the relative Ca, Mg and Fe abundances of a real sample of 202 cool DZs, identified from SDSS data by \citet{Hollands2017}, can be reproduced reasonably well by an underlying fragment core fraction distribution which is a Delta function, combined with random Gaussian noise of roughly 0.1 to 0.25 dex. Physically, this corresponds to a model in which accreted material is either undifferentiated, or has experienced minimal fragmentation following differentiation. Apparent enhancements (or depletions) of Fe in individual systems are therefore due to random noise, rather than being indicative of geological processes.

While the best model (Delta) invokes accretion of undifferentiated material, the data can also be matched by the Collisional model, in which white dwarf pollutants are fragments from collisions. This requires the random noise to be below 0.15 dex. In this model, the spread in pollutant compositions is partially due to random noise, but is also partially due to genuine variation in the core fractions of accreted pollutants across the sample. However, Figure~\ref{fig:SyntheticHollandsCramer} indicates that this model is only borderline acceptable, and should be disfavoured compared to the Delta model.

The higher noise level implied by the Delta model, compared to the Collisional model, better matches the typical errors on the real data (which are roughly 0.1 to 0.2 dex). Indeed, on the assumption that the Delta model is accurate, the implied noise range of 0.1 to 0.25 dex acts as an independent test of the accuracy of the real error estimates. In this case, our results broadly agree with these estimates. The allowed error range in the Collisional scenario ($<0.15$ dex) aligns less neatly with the data. However, it is plausible that the errors on the real data are overestimated due to correlations between abundances, which we do not take into account (see Section~\ref{sec:errors}).

Overall, we consider the Delta $f_{\rm c}$ distribution to be the best supported by the modelling considered here. This model implies that accreted material is primitive, and any apparent evidence of differentiation is due to random noise. However, an analysis of the same sample of 202 cool DZs at the individual system level finds 7 systems\footnote{These systems are SDSS J0010-0430, SDSS J0744+4649, SDSS J0823+0546, SDSS J0939+4136, SDSS J1043+3516, SDSS J1234+5208 and SDSS J1340+2702.} with strong ($>3\sigma$) evidence of core-mantle differentiation \citep{Harrison2021,BuchanThesis}. This is highly unlikely to occur by chance (0.016\%). All 7 systems feature additional corroborative elements not considered here (Ni or Cr)\footnote{For SDSS J0744+4649 only, Na is also a corroborative element because this system was inferred to be accreting crustal material, of which Na is diagnostic.} and/or atypically small abundance uncertainties, such that it is more difficult to explain these systems by random noise. When the corroborative elements are removed, and errors are increased to 0.2 dex where appropriate, only SDSS J0823+0546 is still significant at the $3\sigma$ level. The probability of randomly finding at least one $3\sigma$ detection from a sample of 202 is 42\%. In other words, if these systems are treated in a way which is more compatible with our three-element, single-valued random noise model, they would not appear to be so anomalous.

Nevertheless, these systems do demonstrate that the undifferentiated (Delta) model cannot be the full story. However, we are unable to place constraints on any differentiated component of the pollutant population given the available sample size and the errors on the abundances.

A better model may invoke a mixture of the three evolutionary scenarios considered here, and possibly others. However, such a mixture model introduces additional complexity. A larger sample, smaller abundance uncertainties and/or more elements would be needed to tease out these additional refinements. 22 of the cool DZs feature detections of the siderophile element Cr, which can be used as a diagnostic for $f_{\rm c}$. However, we find that this subset is not large enough to be statistically inconsistent with any of the three accretion models (at any noise level), despite the additional sensitivity provided by the inclusion of this extra element in the analysis.

Some members of the cool DZ sample with clear Fe lines received follow-up observations, leading to higher signal-to-noise ratios. Therefore, such a mixture model should consider the possibility of systematically lower noise on systems with high Fe abundance.

\subsection{Sources of core-rich bias}

The bias towards preferential detection of core-rich material is a direct result of the relative ease of detecting Ca relative to Fe, as noted by \citet{Bonsor2020}. This compounds an additional core-rich bias which can be introduced during modelling of relative metal abundances. Mantle-rich material and differential sinking are typically degenerate, as both processes deplete Fe relative to Ca and Mg. Within the Bayesian frameworks of \citet{Harrison2021}, \citet{Buchan2021} and \citet{Bonsor2023}, this causes a bias against inference of mantle-rich material as this requires an extra free parameter, which is strongly penalised. Indeed, these papers find that core-rich systems heavily outnumber mantle-rich systems. A similar degeneracy is present in the Bayesian analysis of \citet{Swan2023a}, although the presence of (at least) 5 elements in their data, along with several upper bounds, reduced any resulting bias. Future work aiming to calculate the relative frequency of core- and mantle-rich pollutants should take these biases into account. The detection of multiple elements (or estimation of upper bounds) can mitigate against degeneracy.

\subsection{Implications for future surveys}
\label{sec:future_surveys}

This work highlights the need for unbiased surveys in the future, or at least surveys with well-understood biases. If we are to use an observed sample of polluted white dwarfs to interpret the true distribution of composition in the accreted bodies, it is key to know how the sample was selected and the sensitivity to detection of different elements in each white dwarf. This work finds that in order to have a good ($\gtrsim80\%$) chance at drawing conclusions on whether a given $f_{\rm c}$ distribution can explain compositional trends, we require a sample size on the order of hundreds, with 500 white dwarfs almost guaranteeing statistical discrimination in our three cases. This assumes that all white dwarfs have detections of multiple elements, whose abundances have typical errors of 0.2 dex.

Figure~\ref{fig:dprobheatmap} may be a useful guide for how these effects should be weighed against each other, if the expected error on abundances can be estimated. Note that the sample size on the horizontal axis includes only those systems for which $f_{\rm c}$ can be estimated, which requires detection of multiple elements. For H-dominated systems, estimating $f_{\rm c}$ requires at least one siderophile and one lithophile, plus another element to break the degeneracy on nebular composition. For He-dominated systems, the number of elements required is more difficult to determine because retrieval of $f_{\rm c}$ was not possible with our simple model. However, Section~\ref{sec:hollands} demonstrated that direct comparison of three elements (Ca, Mg and Fe) may be sufficient.

The additional complexity of modelling the pollutants accreted by He-dominated systems is ultimately due to their comparatively long sinking timescales: from a modelling perspective, shorter is better. Higher values of ${\rm T}_{\textrm{eff}}$ and $\log(g)$ are therefore also favourable in general. However, there is no reason why different types of white dwarf (i.e., H- and He-dominated) cannot be combined into a larger sample and modelled together as in Section~\ref{sec:hollands}.

We do not consider the effect of resolution on our detection thresholds explicitly. This is because the resolution should also affect the typical observational error, and initial testing showed that the error effect was dominant. In Figure~\ref{fig:dprobheatmap}, the effect of resolution is therefore indirectly tied to the error shown on the vertical axis.

\subsection{The importance of collisions in exoplanetary systems}

White dwarfs accrete planetary bodies of at least tens of kms \citep{Jura2003}. The study of main sequence debris discs leaves an open question regarding whether such large planetesimals are collisionally evolved. These debris discs are detected via infrared emission, which arises from small dust grains that are continually replenished via a collisional cascade. The size of the largest bodies which feed the cascade is unclear. Until recently, the largest debris was often assumed to be dwarf-planet sized, like in our Solar System, but this cannot be true for the largest, brightest debris discs, otherwise their mass would exceed that available in protoplanetary discs. This problem is resolved if the maximum size of bodies which participate in the collisional cascade is on the order of a few km \citep{Krivov2021}. White dwarf observations could act as a test of this theory, if they were to provide evidence as to whether larger planetesimals are collisionally evolved and therefore part of a collisional cascade.

\subsection{Limitations and caveats of our analysis}

The most important limitations of the cool DZ analysis are the uncertainties on our relative sinking timescales for the He-dominated systems, and the possibility that the initial nebular composition could differ from our assumed values. For the analysis of the H-dominated systems, the biggest caveat is that our choice of significance threshold (used to determine whether two distributions are distinct) is arbitrary. We focus on these issues first, before discussing other caveats which have less impact on our results.

\subsubsection{Relative Sinking Timescales}

Throughout this work, we must consider the effects of differential sinking, which requires knowledge of the relative sinking timescales for key metals. Inaccuracies in these timescales could significantly affect our predicted synthetic abundances for the cool DZ sample. We use sinking timescales from \citet{Koester2020}, with convective overshoot included and $\log(\textrm{Ca}/\textrm{He})$ fixed to $-9.5$ where relevant. The uncertainty in these timescales can be estimated by comparison against timescales taken from the Montreal White Dwarf Database, following \citet{Fontaine2015a,Fontaine2015b}. These sources use independent prescriptions for the primary model components: collision integrals, average ionic charges, and atmosphere models/envelope integration. For He-dominated systems, the MWDD timescales neglect the effect of metals, so for the sake of making a fair comparison we reduce our assumed value of $\log(\textrm{Ca}/\textrm{He})$ to the minimum value of -15; we find this makes little difference to relative sinking timescales, however. We use this approach to estimate the uncertainty in $\tau_{\rm Ca}/\tau_{\rm Fe}$, $\tau_{\rm Ca}/\tau_{\rm Mg}$ and $\tau_{\rm Fe}/\tau_{\rm Mg}$. We did not explore the full temperature range of the cool DZ sample (the MWDD does not make predictions for He-dominated white dwarfs below 7000\;K).

For H-dominated white dwarfs, we find that $\tau_{\rm Ca}/\tau_{\rm Mg}$ and $\tau_{\rm Fe}/\tau_{\rm Mg}$ can be significantly discrepant (by up to about 30\% for the most relevant region of parameter space). We anticipate that this discrepancy should not have a major impact on our results for the H-dominated systems because it affects all our synthetic populations in a uniform manner.

The effect on He-dominated systems is more important, with the impact on $\tau_{\rm Ca}/\tau_{\rm Fe}$ being especially strong. Our timescales predict that $\tau_{\rm Ca}/\tau_{\rm Fe} \sim 1.4$ to within a few percent (with or without convective overshoot), while MWDD predicts $\tau_{\rm Ca}/\tau_{\rm Fe} \sim 1$, with $\tau_{\rm Fe}$ sometimes being greater than $\tau_{\rm Ca}$. This would imply that differential sinking does not strongly affect the observed Ca/Fe ratio, while our adopted timescales predict that Ca/Fe could increase by a factor of up to 4 for the extreme case of observation 5 sinking timescales after the end of accretion. Comparison to Figure~\ref{fig:ternaryplots} suggests that this uncertainty is therefore large enough to potentially change which of our models best matches the cool DZ sample.

These discrepancies strongly motivate future work to explore the impact of different treatments of differential sinking more fully.

\subsubsection{Stellar composition}

The progenitors of the cool DZ stars may have systematically different compositions from the FGK stars assumed in Section~\ref{sec:initialcomp} because stellar composition evolves over galactic time (e.g., \citealt{Ness2019}), and cool white dwarfs must have formed at early times (if not, they would still be warm). Data extracted from the Hypatia catalog \citep{Hinkel2014} implies that Ca/Fe or Mg/Fe could be higher by roughly 0.1 or 0.2 dex (a factor of 1.5) for such stars. Figure~\ref{fig:ternaryplots} implies that a shift of this magnitude could significantly change our predicted synthetic compositions.

\subsubsection{Significance thresholds}

Our estimates of the probability of distinguishing two distributions (Figures \ref{fig:fcf_comparisons} and \ref{fig:dprobheatmap}) depend on the choice of significance threshold. We deem two distributions to be different if the p-value of a KS test is less than 0.05, but this number is arbitrary. Table~\ref{tab:significance_thresholds} exemplifies how different choices propagate into our sample size estimates, showing that different typical choices can change our estimates by a factor of roughly 2.

\begin{table}
    \centering
\caption{Approximate sample size required to distinguish between two populations of H-dominated systems with 90\% probability, as a function of the chosen significance threshold. We use 0.05 throughout this work. The two populations are assumed to have 0.2 dex errors on all abundances, and are generated according to the specified $f_{\rm c}$ distributions. Collisional is abbreviated to `Col'. Orbit-by-orbit is abbreviated to `Obo'. The populations are distinguished (or not) by comparing the retrieved $f_{\rm c}$ distributions. Sample sizes include only systems for which $f_{\rm c}$ can be estimated, which requires detection of multiple elements.}
\label{tab:significance_thresholds}
    \begin{tabular}{cccc}
         \hline
         Threshold& Delta/Col &  Delta/Obo & Obo/Col\\
         \hline
         0.1& 200 & 75 & 225\\
         0.05& 275 & 100 & 275 \\
         0.01& 300 & 125 & 350\\
         0.005& 400 & 150 & 450 \\
         \hline \\
    \end{tabular}

\end{table}

\subsubsection{Robustness of detection thresholds}
\label{sec:detectionthresholdoffsets}

Our results from the synthetic H-dominated systems are sensitive to our calculation of detection thresholds (see Section~\ref{sec:detectionthresholds}), which result in a higher detection rate of lithophiles (Ca, Mg) than siderophiles (Fe). This leads to the inference of core-rich bias and affects the proportion of a given polluted white dwarf sample for which $f_{\rm c}$ can be estimated. Qualitatively, we expect this result to be robust in general because the strength of Ca (and even Mg) lines is typically high compared to Fe lines (e.g., \citealt{Klein2010,Xu2014}). However, if one were to observe in a wavelength range in which the strongest lines are associated with siderophile elements such as Fe, this bias would be completely reversed.

In the synthetic populations, we generally find that Mg is the most commonly detected element, followed by Ca, which is unexpected (in reality, Ca is the most frequently detected element) but doesn't affect this conclusion because both Mg and Ca are lithophiles. This may well be attributable to the detection thresholds. Figure~\ref{fig:detection_thresholds} shows that the detection thresholds for Mg are generally lower than real detected values. This suggests that the detection thresholds are overly lenient on Mg, and mock detections are sometimes made which would not realistically be possible.

Throughout this work, we vary noise level freely while using one set of detection thresholds, while in reality they should be linked. Higher resolution leads to less restrictive thresholds and also to smaller abundance errors. Modelling this effect is beyond the scope of this work.

\subsubsection{Time-averaging and accretion of multiple bodies}

The sinking timescales for He-dominated white dwarfs can be comparable to the timescale of accretion. This means that the pollution present in the atmosphere is an average over some extended period of time, during which the accreted composition may have changed non-negligibly. We ignore this effect, calculating instantaneous pollution compositions (as modified by differential sinking) instead. Time averaging would inevitably lead to a narrower spread of predicted compositions.

It is also possible that white dwarfs accrete material from multiple bodies (with different $f_{\rm c}$) simultaneously \citep{Jura2008,Turner2020,Johnson2022}. In practise, this would have very similar effects to time-averaged accretion, leading to a narrower spread in detected compositions. However, it could affect H-dominated systems as well as He-dominated systems.

Both time-averaging and accretion of multiple bodies could significantly increase the sample sizes needed to distinguish between accretion models, as the extreme $f_{\rm c}$ values which act as diagnostics may get washed out.

However, our result that the Delta model best explains the cool DZ sample would be unchanged, because even in the extreme case that the spread of the Collisional and Orbit-by-orbit distributions is completely averaged away, they reduce to a Delta function. The Orbit-by-orbit distribution in particular ought to be averaged over multiple orbits, as the timescale for compositional variation is $\sim$\SI{0.1}{Myr} \citep{Brouwers2023a} which is less than typical sinking timescales. Additionally, differently sized fragments accrete at different rates so if multiple bodies are accreted during the same accretion event, their compositions will average out to some extent. Broadly speaking, the more one models these effects, the more the Orbit-by-orbit distribution would resemble the Delta distribution. Figures \ref{fig:ternaryplots} and \ref{fig:SyntheticHollandsCramer} suggest that this would better match the data, but it would also make the Orbit-by-orbit scenario somewhat redundant.

\subsubsection{The most likely accretion phase}

For He-dominated white dwarfs, the distributions of $t$ and $t_{\rm event}$ assumed in Section~\ref{sec:phaseofaccretion} imply that the majority of systems are in declining phase, with a smaller proportion in build-up/steady state. This qualitatively agrees with the analysis of \citet{BuchanThesis}, but that analysis is ultimately dependent on estimates of how long accretion events typically last.

For simplicity, we assume the choice of $f_{\rm c}$ distribution does not affect the distribution of $t_{\rm event}$ and $\lambda$ (and that they are not correlated with the sampled value of $f_c$). These variables could be linked via a more thorough model.

The sample size requirements for H-dominated white dwarfs may be underestimated by up to $\approx20\%$. This is because we have assumed that all such systems are in a steady state of accretion, both during synthesis and parameter retrieval. This is a safe assumption unless the white dwarf is particularly cool and does not have high $\log(g)$. In this case, typical sinking timescales can be comparable to the accretion event timescale, which we assume, for the purpose of this calculation, is typically on the order of \SI{1}{Myr} (as in \citealt{Cunningham2021}). The probability of a H-dominated system not being in steady state is then significant if the Mg sinking timescale for that system is greater than roughly $10^5$ yr. We find that 190 of a random sample of 1000 synthetic H-dominated systems meet this criterion. In a worst case scenario the offending systems could be removed from the sample, although a more sophisticated model would be able to factor in the longer sinking timescales anyway.

\subsubsection{Errors}
\label{sec:errors}

We assume that all abundance errors are Gaussian and symmetric (in log space). This replicates how abundances are typically reported, but it is worth noting that asymmetric errors could potentially mimic $f_{\rm c}$ distributions which skew in a particular direction. We assume that errors are equal for each element within a system, and equal for every system within a sample. The former assumption seems justifiable based on the cool DZ sample, but the latter assumption is more questionable: a system with low signal-to-noise ratio is likely to have large errors on all elements. This could potentially be modelled by varying the noise level within a sample according to a Gaussian distribution, then sampling from that noise level according to a Gaussian distribution. However, since the convolution of two Gaussians is another Gaussian, this should be equivalent to our current treatment.

We also assume errors associated with different elements in the same system are uncorrelated. This is unlikely to be true, since abundance uncertainties are partially dependent on the uncertainties on stellar parameters introduced during atmospheric modelling, which are common to all elements (see the Appendix of \citealt{Klein2021} for further discussion). However, in the absence of covariance matrices describing this effect, it is difficult to assess its impact. Our estimate of the noise level for the DZ sample (roughly 0.2 dex) may be an underestimate, since we effectively ignore any correlated component. The errors found by \citet{Hollands2017} include any correlated component which may be present. We also neglect systematic sources of error such as the optical-UV discrepancy \citep{Xu2019}, but since all element abundances for the cool DZ sample are estimated from optical data this discrepancy is not an issue here.

\subsubsection{Core and mantle composition}

We assume that the composition of the core and mantle components of any pollutant body is similar to that of Earth, but scaled according to its initial nebular composition. However, white dwarf pollutants may instead derive from reservoirs more analogous to meteorites or Mars. The primary compositional trend within the Solar System is depletion of volatile elements. Since Ca, Mg and Fe are not volatile, we expect that scaling to the bulk composition of other Solar System reservoirs should not affect our results much. In this sense, our Delta model could be interpreted as the accretion of chondritic material.

Adjusting the core and mantle composition to match those of Mars would likely affect our results more. Figure~\ref{fig:ternaryplots} illustrates how Mars' mantle compositions compares to Earth's (the cores are both 100\% Fe for the purposes of this plot). This figure suggests that adopting a Mars-like composition might have a noticeable effect on the predicted composition of mantle-rich fragments, making them less Fe-poor. This could make the Orbit-by-orbit model more viable, but otherwise appears unlikely to have a strong effect on our results.

\subsubsection{Fragment core number fraction distributions}
\label{sec:tidalcnf}

The Orbit-by-orbit distribution used in Section~\ref{sec:fcf_comparison} is a simplification calculated assuming an isolated tidal disruption event, followed by asynchronous accretion. Realistically, there may also be collisions between the resulting fragments. The Collisional and Orbit-by-orbit distributions we present should be viewed as extreme cases. The true $f_{\rm c}$ distribution for a real sample may lie somewhere in between the extremes, in which the sample size estimates in Section~\ref{sec:fcf_comparison} should be viewed as lower bounds. 
We also note that even if collisions and orbit-by-orbit accretion are actually taking place, but the bodies being accreted are undifferentiated, this is observationally indistinguishable from the Delta model.

In the Orbit-by-orbit case, we neglect the temporal evolution of the $f_{\rm c}$ distribution. In the model of \citet{Brouwers2023a}, there is a short burst of core-rich accretion (with very high accretion rate), followed by a very long period in which mantle-rich material accretes at a very slow rate. In our model, a lower accretion rate corresponds to a lower pollution level (all else being equal), indicating a lower chance of detecting mantle-rich material. However, the longer timescale of mantle-rich accretion increases the number of white dwarfs which would be in this state. It is unclear how these effects would trade off, but it is possible that much of the mantle-rich peak (see Figure~\ref{fig:fcnf_dists}) would be undetectable. This might make it harder to distinguish it from the Collisional and Delta cases, increasing the necessary sample sizes. It might, however, allow for a better match to the cool DZ sample.

\subsubsection{Upper bounds}
\label{sec:upperbounds}

We have ignored the possibility of upper bounds on elemental abundances, partially due to their significant additional complexity. However, upper bounds on key elements are potentially a valuable source of information. For example, if Ca was detected and Fe was not, but there was a (low) upper limit on Fe, one could still infer that this material is mantle-rich. The calculation of such upper bounds would help combat the bias towards core-rich systems described in Section~\ref{sec:controlrealistic}.

\section{Conclusions}
\label{sec:synthetic_conclusions}

Observations of polluted white dwarfs have the potential to reveal the geological history of exoplanetary systems. A key diagnostic is the abundance of Fe relative to lithophiles such as Ca and Mg, which can be used to infer accretion of core- or mantle-like material. However, measured abundances can be altered by random noise plus systematic errors and skewed by observational bias, motivating a population level analysis in which large sample sizes overcome these obstacles. We focus on the potential for populations of polluted white dwarfs to discriminate between three models describing the evolution of rocky bodies. The first model is the accretion of highly processed
collision fragments of larger planetary bodies with iron cores. The second model is that the separation of core and mantle material is a natural consequence of orbit-by-orbit accretion. The third model assumes minimal differentiation or, equivalently from an observational perspective, minimal fragmentation. We use synthetic populations of white dwarfs to investigate the conditions under which these models can be disentangled, and the effects of observational bias and random noise.

We find that samples of white dwarfs are likely biased by selection effects towards core-rich material. An additional bias is introduced by random noise, which leads to the preferential (mis)identification of extremely core-rich or mantle-rich material. Taking these biases into account, we calculate the probability that the three accretion models can be distinguished, finding that it increases as sample size increases and as the random noise associated with metal abundances decreases. Assuming 0.2 dex errors, this probability approaches 100\% once the sample size is above 500. This sample size counts only systems with enough detected elements to estimate the fraction of core material in the pollutant.

We apply our techniques to the sample of 202 cool DZs presented by \citet{Hollands2017}, testing whether their relative Ca, Mg and Fe abundances are consistent with those predicted by our accretion models. We find that these abundances are best reproduced, and are reproduced well, by a model in which differentiation and/or fragmentation has not occurred and the apparent spread of compositions is due to random noise. We estimate that the typical error due to noise is roughly 0.2 dex, consistent with the actual reported errors.

However, some individual systems do show convincing evidence for differentiation due to additional corroborative elements and/or particularly small errors on their abundance estimates. Our simple model cannot explain these systems and is therefore incomplete. In order to explain them, a more complete model which additionally invokes a contribution from differentiated material is necessary, but additional data are required to constrain such a model. We also note that our interpretation is sensitive to assumptions about the provenance of pollutants and time elapsed since their accretion onto a white dwarf. Nevertheless, our results illustrate the potential for random noise to mimic geological processes.

Our models make a number of assumptions. Our conclusions for the cool DZ sample are particularly sensitive to the relative timescales on which different metals diffuse through the white dwarf atmosphere. Our conclusions could also be modified if a different initial stellar composition, compared to the progenitors of other white dwarfs, were to be assumed. The prediction for the number of H-dominated white dwarfs needed to distinguish between evolutionary models is sensitive to our arbitrary choice of significance threshold.

Alongside the DESI survey, which has so far discovered 121 white dwarfs with metal lines \citep{Manser2024DESI}, the 4MOST, SDSS-V and WEAVE-WD surveys will dramatically expand the population of known polluted white dwarfs (the WEAVE-WD survey aims to provide spectroscopy from 100,000 systems). The \textit{Gaia} data release DR3 \citep{Gaia2023} includes a sample of 100,000 white dwarfs, which may contain systems with as-yet undiscovered metal pollution. \citet{Vincent2023} identified 896 high-confidence DZs within this sample, and \citet{Kao2024} used machine learning to find 375 white dwarfs with at least 3 metals. Overall, the prospects for obtaining a sample of a few hundred suitable DZs within a few years appears promising. At the same time, modelling techniques for estimating metal abundances are continually improving. The combination of these effects should dramatically improve the potential for populations of white dwarfs to reveal the key processes which govern the evolution of planetary systems.

\section*{Acknowledgements}

AMB is grateful for the support of a PhD studentship funded by a Royal Society Enhancement Award,  RGF\textbackslash EA\textbackslash 180174. This research was additionally supported by a Leverhulme Trust Grant (ID RPG-2020-366). AB acknowledges the support of a Royal Society University Research Fellowship, URF\textbackslash R1\textbackslash 211421. LKR acknowledges support of a Royal Society University Research Fellowship, URF\textbackslash R1\textbackslash 211421 and an ESA Co-Sponsored Research Agreement No. 4000138341/22/NL/GLC/my = Tracing the Geology of Exoplanets. We thank Boris Gaensicke, Mark Hollands, Chris Manser, Detlev Koester, Andrew Swan, Matthew Auger-Williams and Tim Pearce for helpful discussions and advice. We also thank the anonymous referee for providing useful feedback which improved the clarity of the manuscript. The research shown here acknowledges use of the Hypatia Catalog Database, an online compilation of stellar abundance data as described in \citet{Hinkel2014}, which was supported by NASA's Nexus for Exoplanet System Science (NExSS) research coordination network and the Vanderbilt Initiative in Data-Intensive Astrophysics (VIDA).

\section*{Data Availability}
\label{sec:data}

The data and code used in this work are publicly available at \url{https://github.com/andrewmbuchan4/PyllutedWD_Public}.




\bibliographystyle{mnras}
\bibliography{references.bib}



\appendix

\section{Detection thresholds}

\begin{figure*}
    \centering
    \includegraphics[width=\textwidth,keepaspectratio=true]{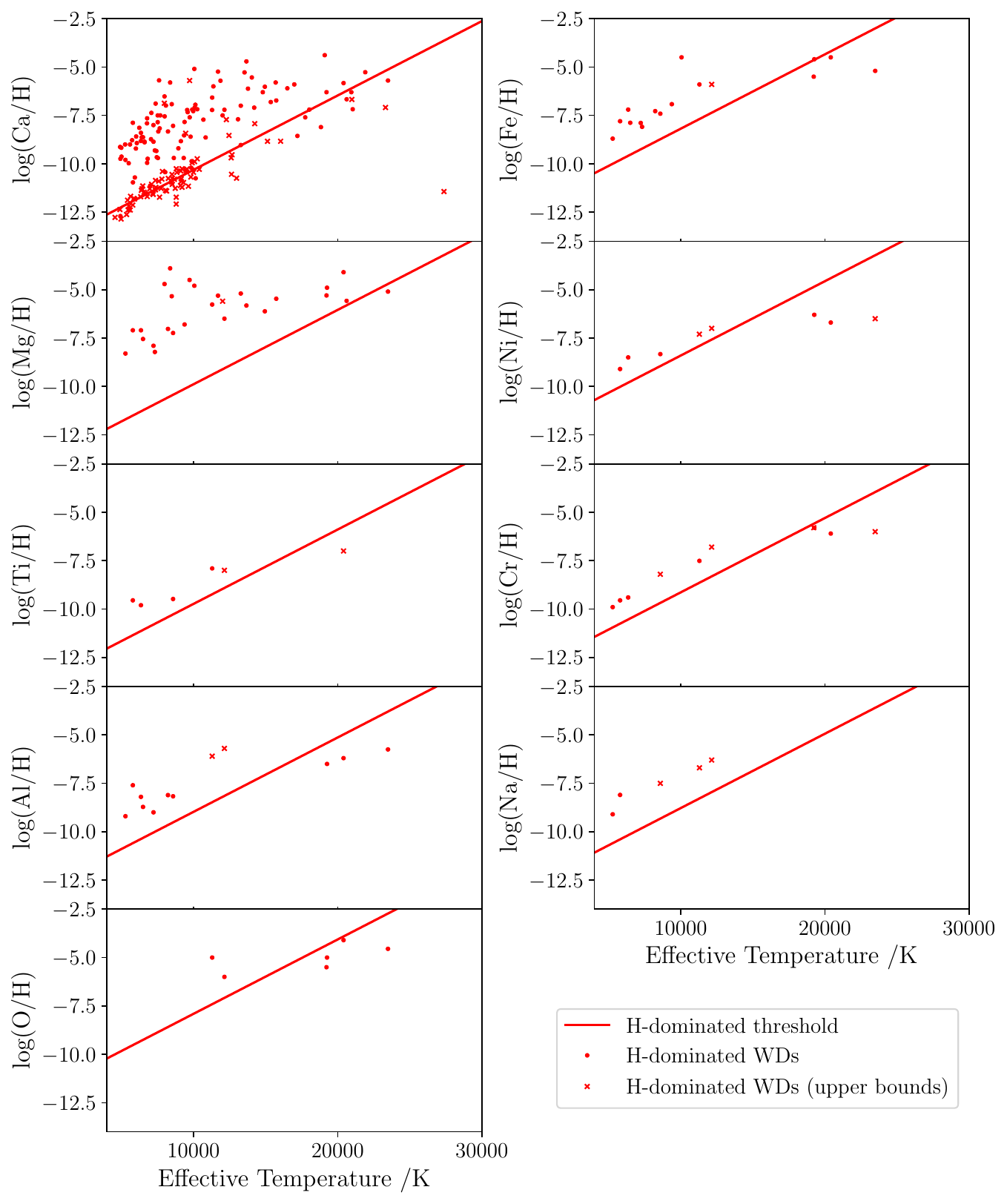}
    \caption{Illustration of the detection thresholds for H-dominated white dwarfs for each element considered in this work. Abundances above the relevant solid line for the specified element at a certain temperature are deemed detectable. The temperature range shown here spans the full range we consider. The calculation of these thresholds is described in Section~\ref{sec:detectionthresholds}. As a sanity check, we also illustrate detected metal abundances (red dots) and upper bounds (red crosses) for real DAs. These data were obtained by exporting all white dwarfs from the Montreal White Dwarf Database (\citealt{MWDD}, accessed 01/08/2022), and compiling all valid abundances from the output. These values provide a relevant comparison, although alternative values may be available.}
    \label{fig:detection_thresholds}
\end{figure*}

\begin{figure*}
    \centering
    \includegraphics[width=\textwidth,keepaspectratio=true]{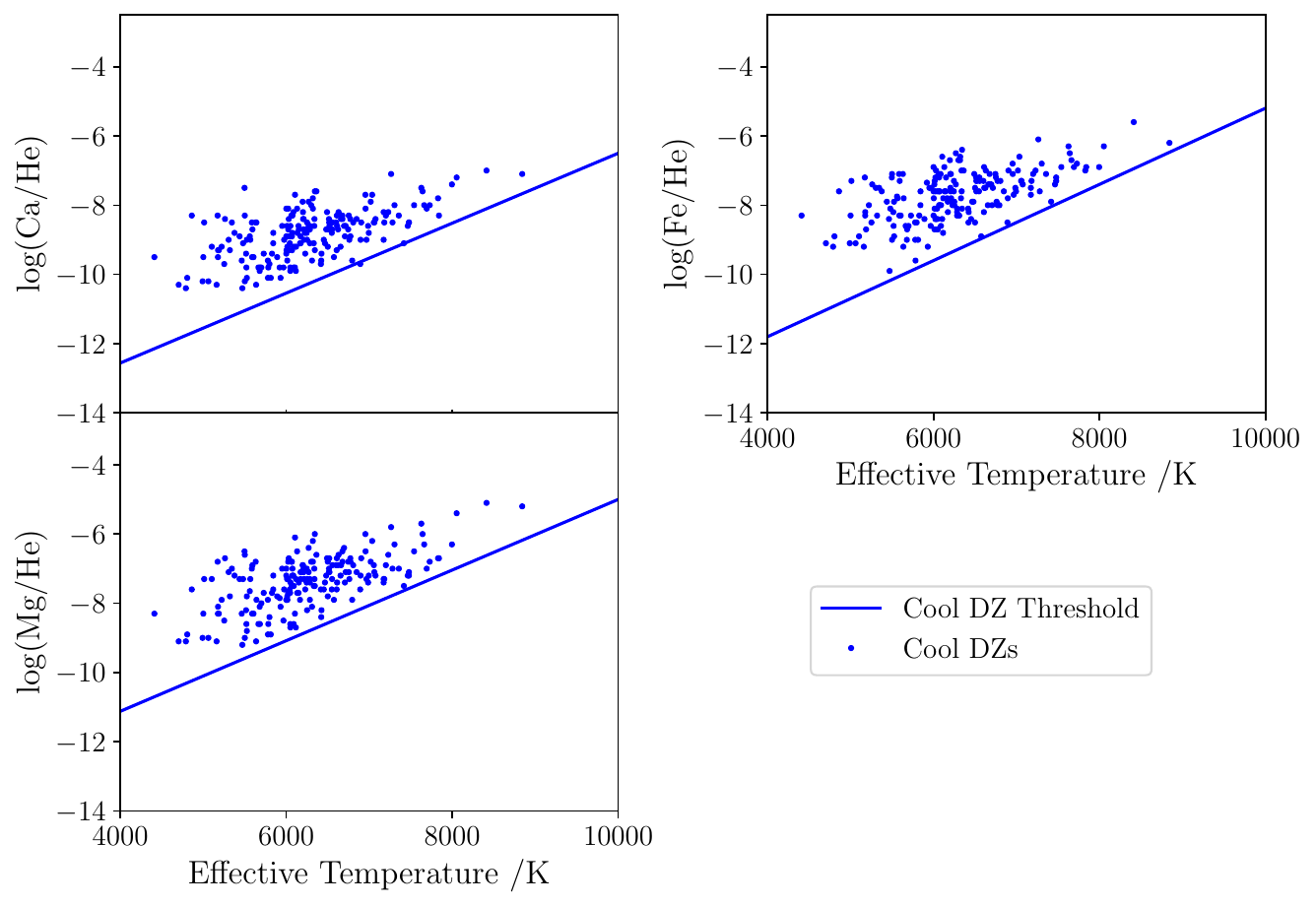}
    \caption{Illustration of the detection thresholds for He-dominated white dwarfs for Ca, Mg and Fe. Abundances above the relevant line for the specified element at a certain temperature are deemed detectable. The temperature range shown here spans the range of the cool DZ sample from \citet{Hollands2017}. The abundances of elements in this sample are also illustrated. The detection thresholds were obtained by manual calibration to these abundances.}
    \label{fig:detection_thresholds_dz}
\end{figure*}

This Appendix includes two Figures (\ref{fig:detection_thresholds} and \ref{fig:detection_thresholds_dz}) illustrating the detection thresholds used.

\bsp	
\label{lastpage}
\end{document}